\documentclass[aps, prd, superscriptaddress, preprintnumbers, showpacs, notitlepage]{revtex4-1}

\usepackage{graphicx,amsfonts,amsmath,amssymb,amstext}
\usepackage{float,wrapfig}
\usepackage{dsfont}

\usepackage{color}
\usepackage{bm}

\usepackage{multirow}

\begin{document}

\title{Generalized Landau level representation: Effect of static screening in the quantum Hall effect in graphene}

\author{Igor A. Shovkovy}
\email{igor.shovkovy@asu.edu}
\affiliation{College of Letters and Sciences, Arizona State University, Mesa, Arizona 85212, USA}

\author{Lifang Xia}
\email{lifang.xia@asu.edu} 
\affiliation{Department of Physics, Arizona State University, Tempe, Arizona 85287, USA}

\date{\today}

\begin{abstract}
By making use of the generalized Landau level representation (GLLR) for the quasiparticle 
propagator, we study the effect of screening on the properties of the quantum Hall states with 
integer filling factors in graphene. The analysis is performed in the low-energy Dirac model in the 
mean-field approximation, in which the long-range Coulomb interaction is modified by the 
one-loop static screening effects in the presence of a background magnetic field. By utilizing a 
rather general ansatz for the propagator, in which all dynamical parameters are running functions 
of the Landau level index $n$, we derive a self-consistent set of the Schwinger-Dyson (gap) 
equations and solve them numerically. The explicit solutions demonstrate that static screening 
leads to a substantial suppression of the gap parameters in the quantum Hall states with a broken 
$U(4)$ flavor symmetry. The temperature dependence of the energy gaps is also studied. 
The corresponding results mimic well the temperature dependence of the activation energies 
measured in experiment. It is also argued that, in principle, the Landau level running of the 
quasiparticle dynamical parameters could be measured via optical studies of the integer 
quantum Hall states.
\end{abstract}

% PACS, the Physics and Astronomy Classification Scheme.

%\keywords{graphene, quantum Hall effect, Coulomb interaction, symmetry breaking, static screenig}
%Use showkeys class option if keyword display desired

\maketitle

\section{Introduction}

As predicted theoretically more than three decades ago \cite{Semenoff:1984dq,DiVincenzoMele:1984}, 
the low-energy excitations in planar graphite (also known as graphene) are described by $(2+1)$-dimensional, 
massless Dirac fermions. Interestingly, the spinor structure of the corresponding fermion fields has 
nothing to do with the usual spin of electrons. It is connected with an effective ``pseudospin" that 
has its roots in the hexagonal arrangement of carbon atoms in graphene, which can be viewed 
as a superposition of two inequivalent triangular sublattices. At the same time, the usual spin 
plays a rather passive role analogous to an extra ``species" degree of freedom. 

The Dirac nature of excitations was confirmed experimentally by the observation of the anomalous 
integer quantum Hall (QH) effect in Refs.~\cite{novoselov:2005ntr,zhang:2005ntr}. At sufficiently low
temperatures, the measured Hall conductivity revealed a set of well-resolved QH plateaus at the 
filling factors $\nu=\pm4(k+1/2)$, where $k=0,1,2,...$ is an integer. One of the signature properties 
of the corresponding QH effect is the anomalous shift of $1/2$ in the expression for the filling factor. 
This measurement appears to be in perfect agreement with the theoretical predictions of the 
low-energy Dirac theory \cite{Gorbar:2002iw,zheng:2002prb,Gusynin:2005pk,peres:2006prb}. 
The overall factor of $4$ in the expression for $\nu$ is also in agreement with the predicted fourfold 
(spin and sublattice-valley) degeneracy of the Landau levels in the low-energy theory. For 
reviews of quantum Hall physics in graphene see Ref.~\cite{CastroNeto:2009zz,Chakraborty2010,Barlas2012}.

The spin and sublattice-valley degeneracy of the Landau levels in the low-energy theory of 
graphene can be associated with an approximate global ``flavor" $U(4)$ symmetry. Strictly 
speaking, the symmetry is not exact. There exist several small explicit symmetry breaking effects 
that lift the degeneracy of the Landau levels \cite{Aleiner:2007va}. The most obvious among 
them is the Zeeman effect that breaks the $U(4)$ symmetry down to $U_{\uparrow}(2)
\times U_{\downarrow}(2)$, where $U_{s}(2)$ with $s=\uparrow,\downarrow$ is the sublattice-valley 
symmetry of quasiparticles with a fixed spin. In practice, however, the flavor $U(4)$ symmetry 
could be treated as exact because, for any realistic value of the magnetic field, the Zeeman 
energy is much smaller than the Landau energy scale. Moreover, the Zeeman energy, as 
well as all other explicit symmetry breaking effects are negligible not only compared to the 
Landau energy scale, but even compared to the nonzero (thermal/interaction) widths of the 
individual Landau levels. This explains why it is hard to resolve experimentally the QH 
plateaus with any filling factors other than $\nu=\pm4(k+1/2)$ that correspond to fully filled 
fourfold-degenerate Landau levels. 

The subsequent experimental studies revealed that, in sufficiently strong magnetic fields, 
additional QH plateaus at $\nu=0,\pm1,\pm3,\pm4$ can be resolved \cite{zhang:2006prl,Abanin2007,jiang1:2007prl,Checkelsky2008,Giesbers2009,Checkelsky2009,
du:2009ntr,bolotin:2009ntr,Zhang2010,Zhao2012,Young2012,Young2014,Chiappini2015}.
As is clear, the corresponding 
QH states have fractional fillings of the low-lying Landau levels. The implication of this is that the 
approximate fourfold degeneracy of the Landau levels is truly lifted. Such new QH states could be 
explained theoretically if the electron-electron interaction triggers spontaneous breaking of 
the flavor $U(4)$ symmetry. For example, the symmetry breaking $U(4)\to U_{\uparrow}(2)
\times U_{\downarrow}(2)$ by a {\em dynamically enhanced} Zeeman splitting could potentially 
explain the origin of the QH states with $\nu=0$ and $\nu=\pm 4$. If realized, such a scenario
would be nothing else, but a textbook example of the QH ferromagnetism (QHF) 
\cite{nomura:2006prl,abanin:2006prl,goerbig:2006prb,alicea:2006prb,sheng:2007prl,
semenoff:2011jhep} [see also Ref.~\cite{fogler:1995prb}].  

It should be emphasized, however, that QHF is not the only possibility. There exist a number 
of symmetry breaking mechanisms and residual symmetries consistent with the filling factors of 
the additional QH states. One of such alternative scenarios utilizes the idea of magnetic catalysis (MC) 
\cite{Gusynin:1994re,Shovkovy:2012zn,Miransky:2015ava}. The corresponding order parameters are 
excitonic (particle-hole) condensates responsible for the generation of the Dirac and/or Haldane 
masses of quasiparticles \cite{Gorbar:2002iw,Khveshchenko:2001zz,gusynin:2006prb,fuchs:2007prl,
ezawa:2007jpsj,gorbar:2008ltp,gorbar:2008prb,herbut:2006prl,HerbutRoy}. At the microscopic level, 
the excitonic condensates corresponds to a charge density wave (CDW), or a valley polarized CDW. 
Symmetry arguments, as well as direct effective model studies \cite{gorbar:2008ltp,gorbar:2008prb} 
suggest that the order parameters associated with the MC and QHF scenarios necessarily coexist. 
Also, in principle, these two could reproduce all integer QH plateaus observed experimentally in 
strong magnetic fields. With that being said, a number of different types of order parameters are 
possible \cite{herbut:2006prl,HerbutRoy,Kharitonov,Jung-MacDonald2009,Nomura-2009,Roy-DasSarma}. 
It is also fair to note that the precise nature of the observed integer QH states is not always 
unambiguous from theoretical considerations and not always established unambiguously in 
experiment. 

From the experimental point of view, a lot of effort was devoted to revealing the underlying nature 
of the strongly insulating $\nu=0$ QH state, associated with half-filling of the lowest Landau level
\cite{zhang:2006prl,Abanin2007,jiang1:2007prl,Checkelsky2008,Giesbers2009,Checkelsky2009,
du:2009ntr,bolotin:2009ntr,Zhang2010,Zhao2012,Young2012,Young2014,Chiappini2015}. The
main advances in resolving the nature of the QH states were made by applying a tilted magnetic 
field to high-quality graphene devices, that were fabricated on a thin hexagonal boron nitride 
substrate \cite{Young2012,Young2014} or encapsulated between two layers of hexagonal 
boron nitride \cite{Chiappini2015}. In the case of $\nu=0$ QH state, in particular, a careful 
analysis of the conductance and the bulk density of states \cite{Young2012,Young2014,
Chiappini2015} suggests that the state is not spin polarized when the magnetic field is 
perpendicular to the plane of graphene. This is consistent, for example, with both an 
antiferromagnet and a charge density wave. When an in-plane component of the magnetic 
field increases, the state gradually transforms into a fully polarized ferromagnetic state. 
In the intermediate regime, a canted antiferromagnetic state is presumably realized 
\cite{Kharitonov}. Such an interpretation is supported by the observation of the bulk 
gap that does not close and the edge states that become conducting with increasing 
of the total magnetic field. Considering, however, how heavily the arguments rely on 
the properties of the edge states, the final conclusions should be still accepted with 
a caution.

The high-quality graphene devices in Refs.~\cite{Young2012,Chiappini2015} 
also reveal a large sequence of integer QH states with $\nu\geq 1$. Among these, there
are two states associated with quarter filling of the lowest Landau level, i.e., $\nu=\pm1$. 
The linear growth of the energy gaps in these two states as a function of the total magnetic 
field points to their spin polarized nature and the major role played by the Zeeman energy 
in aligning the ground state. A similar sensitivity to the Zeeman energy is also observed 
for the states associated with half-filling of higher Landau levels, i.e., $\nu=4n$, but not 
for the states with quarter and three-quarter fillings, i.e., $\nu=4n\pm 1$ \cite{Young2012}.
As we will see, most of these features are reproduced in the model studied here.

In the mean-field approximation, the role of the long-range Coulomb interaction and 
Landau level mixing were studied in Ref.~\cite{gorbar:2012ps}. By utilizing a rather 
general combination of the MC and QHF order parameters, the corresponding study 
was able to reproduce all observed QH plateaus (i.e., $\nu=0,\pm1,\pm3,\pm4$) as 
well as to suggest that QH plateaus with any integer filling are possible. The symmetry 
breaking patterns of the solutions obtained in the model with the long-range Coulomb 
interaction appeared to be similar to those in the model with local four-fermion 
interaction in Ref.~\cite{gorbar:2008prb}. The qualitative effects due to the 
long-range force were (i) Landau level mixing and (ii) ``running" of all dynamical
parameters (i.e., the wave-function renormalization parameters, the Dirac and Haldane 
masses) as functions of the Landau level index. 

In this paper, we extend the study of the abnormal integer QH effect in the model with the 
long-range Coulomb interaction \cite{gorbar:2012ps} by including the screening effects and 
the effects of nonzero temperature. By making use of a similar mean-field approximation 
that neglects the fluctuations of order parameters, we will derive the full quasiparticle 
propagators in the GLLR formalism for all qualitatively different series of QH states with 
integer filling factors. (A semirigorous justification of the approximation used will be provided 
in Sec.~\ref{SDEscreening}.) As expected, the corresponding results 
contain not only the information about the energy gaps at the Fermi level, but also 
the complete dispersion relations of quasiparticles in all Landau levels. The resulting
propagators can be used in transport calculations, predictions of various emission 
and absorption rates, etc.

This paper is organized as follows. In Sec.~\ref{sec:model}, we introduce the model,
define the quasiparticle propagator, and derive the gap equations. In Sec.~\ref{sec:numerical}, 
we present our numerical analysis of the gap equations and classify the main types of 
solutions. We also compare our results with those in the previous studies. The summary 
of the main results is given in Sec.~\ref{sec:discussion}.

\section{The Model}
\label{sec:model}

Following the same approach as in Ref.~\cite{gorbar:2012ps}, in this study we will use the language of the
low-energy theory for the description of the QH effect in monolayer graphene. The low-energy quasiparticle 
fields are given by the following four-component Dirac spinors 
$\Psi_s = ( \psi_{KAs}, \psi_{KBs}, \psi_{K^\prime Bs}, -\psi_{K^\prime As})$, which combine the Bloch 
states on the two different sublattices $(A,B)$ of the hexagonal graphene lattice in coordinate space. 
The components labeled by the valley indices $K$ and $K^\prime$ correspond to the Bloch states with 
the momenta from the vicinity of the corresponding inequivalent Dirac points ($K$ or $K^\prime$) in
the two-dimensional Brillouin zone. The field $\Psi_s$ also carries an additional index that describes
the spin state, i.e., $s=\uparrow,\downarrow$. 

\subsection{Quasiparticle Hamiltonian}

The free quasiparticle Hamiltonian has the following pseudorelativistic form:
\begin{equation}
H_0 = v_F \int d^2\mathbf{r} \bar{\Psi} (\gamma^1 \pi_x + \gamma^2 \pi_y) \Psi, 
\label{freeHamiltonian}
\end{equation}
where $v_F\approx10^6 \mbox{m/s}$ is the Fermi velocity, $\mathbf{r}=(x, y)$ is the position vector 
in the plane of graphene, and $\bar{\Psi} = \Psi^{\dagger}\gamma^0$ is the Dirac-conjugated 
spinor. Note that the sum over the spin states is implicit in Eq.~(\ref{freeHamiltonian}).
The canonical momentum $\bm{\pi} = -i\hbar\bm{\nabla}   - e\mathbf{A}/c$ includes the vector 
potential $\mathbf{A}$ that corresponds to the magnetic field component $\mathbf{B}_\perp$
orthogonal to the plane of graphene.
The $4\times4$ Dirac matrices $\gamma^\nu$ (with $\nu=0,1,2$) are defined as follows: 
$\gamma^\nu = \tilde{\tau}^3\otimes (\tau^3, i\tau^2, -i\tau^1)$, where the two sets of Pauli matrices 
$\tilde{\tau}$ and $\tau$ act in the valley $(K,K^{\prime} )$ and sublattice $(A,B)$ 
spaces, respectively. They satisfy the usual anticommutation relations $\{\gamma^\mu,\gamma^\nu\}
=2g^{\mu\nu}$, where $g^{\mu\nu} = \mbox{diag}(1,-1,-1)$, and $\mu,\nu=0,1,2$.  

In order to be able to describe QH states with arbitrary filling factors, we should also
allow for a nonzero electron chemical potential $\mu_{\rm e} $. The latter is introduced via an 
additional term, $-\mu_{\rm e} \int d^2\mathbf{r} \Psi^\dagger\Psi$, in the free Hamiltonian in 
Eq.~(\ref{freeHamiltonian}). 
The long range Coulomb interaction is included in the low-energy Hamiltonian by adding 
the following term:
\begin{equation}
H_C = \frac{1}{2} \int d^2\mathbf{r} d^2\mathbf{r}^\prime 
\Psi^{\dagger}(\mathbf{r}) \Psi(\mathbf{r}) U_C(\mathbf{r}-\mathbf{r}^\prime) \Psi^\dagger(\mathbf{r}^\prime)\Psi(\mathbf{r}^\prime),
\end{equation}
where $U_C(\mathbf{r}-\mathbf{r}^\prime)$ is the coordinate-space Coulomb potential in the 
presence of a constant magnetic field. It is straightforward to check that the resulting Hamiltonian 
$H=H_0+H_C-\mu_{\rm e} \int d^2\mathbf{r} \Psi^\dagger\Psi$ is invariant under the flavor 
$U(4)$ symmetry that combines the transformations in spin and sublattice-valley 
spaces. For the explicit form of the symmetry generators see, for example, 
Ref.~\cite{Gorbar:2002iw}. Such a flavor symmetry is explicitly (although weakly) 
broken by the Zeeman interaction of the quasiparticle spin with the external magnetic field 
$\mathbf{B}$. The additional interaction term in the Hamiltonian is given by 
$\int d^2\mathbf{r} \mu_B B \Psi^\dagger \sigma^3 \Psi$, where $B=|\mathbf{B}|$, 
$\mu_B = e\hbar/(2mc)$ is the Bohr magneton, and $\sigma^3$ is the Pauli matrix 
acting in the spin space. As is easy to check, the Zeeman term breaks the $U(4)$ 
symmetry down to the $U_\uparrow(2)\times U_\downarrow(2)$ symmetry. 

The complete model Hamiltonian, including the Zeeman interaction and the electron 
chemical potential $\mu_{\rm e} $, can be conveniently rewritten in the following quasirelativistic 
form:
\begin{equation}
H=\int d^2r \bar{\Psi} \left[ v_F(\bm{\pi}\cdot\bm{\gamma})  - \mu_{\rm e}   \gamma^0 + \mu_B B \sigma^3  \gamma^0 \right] \Psi
+H_C.
\end{equation}
This implies that the inverse bare quasiparticle propagator is defined by the following expression:
\begin{equation}
iS^{-1}(\omega;\mathbf{r},\mathbf{r}^\prime) = \left[ (\omega+\mu_{\rm e}  - \mu_B B \sigma^3 ) \gamma^0  - v_F (\bm{\pi}\cdot\bm{\gamma}) \right] \delta(\mathbf{r}-\mathbf{r}^\prime).
\end{equation}
Here and below, it is convenient to use a mixed $(\omega,\mathbf{r})$-representation. 

Let us note in passing that a certain degree of disorder is always present in real graphene 
devices and, in fact, plays a critical role in the observation of the quantum Hall plateaus. 
However, in the model analysis below, which concentrates primarily on the delocalized 
quasiparticles states in the bulk of graphene, it is justifiable to ignore disorder. Indeed, 
when the spectrum of the delocalized states is established, a large number of bulk 
observables (e.g., symmetry properties of the QH states, density of states, energy of 
transitions lines, activation energies, etc.) will be predicted without much ambiguity. The 
corresponding limitation is not so critical also because of the robustness 
of the QH effect that stems from its topological nature.

\subsection{Dynamical symmetry breaking and full quasiparticle propagator}

In order to be able to describe various QH states with integer fillings factors, we use a
rather general ansatz for the full quasiparticle propagator \cite{gorbar:2012ps},  
\begin{equation}
G^{-1}(\omega;\mathbf{r},\mathbf{r}^\prime) = -i \left[ (\gamma^0\omega - v_F \hat{F}^{+} (\bm{\pi}\cdot\bm{\gamma}) + \hat{\Sigma}^{+} \right] \delta(\mathbf{r}-\mathbf{r}^\prime) ,
\label{fullProp}
\end{equation}
where $\hat{F}^+$ and $\hat\Sigma^+$ are operator-valued wave-function renormalization and self-energy 
functions, respectively. For simplicity, here we will assume that the propagator is a diagonal $2\times 2$ 
matrix in the spinor space, or in other words that the propagator for {\it each} of the two spin states looks
like that in Eq.~(\ref{fullProp}). As is clear, such a choice of the ansatz for the propagator does not 
allow spin mixing. As a result, one cannot describe certain types of states (e.g., a canted antiferromagnetism \cite{Kharitonov}). Nevertheless, we emphasize that the ansatz in Eq.~(\ref{fullProp}) is extremely flexible 
and can describe a large number of states (with very different symmetry properties) for each integer filling 
factor. In the case of filling factor $\nu =0$, for example, these include charge and spin density waves, as well as 
ferromagnetic and antiferromagnetic states. These are basically all main options proposed for a configuration 
with a magnetic field perpendicular to the plane of graphene.

By construction, both $\hat{F}^+$ and $\hat\Sigma^+$ are functions of the three 
mutually commuting dimensionless operators: $\gamma^0$, $is_{\perp}\gamma^1\gamma^2$, 
and $(\bm{\pi}\cdot\bm{\gamma})^2l^2$, where $s_{\perp}=\mbox{sign}(eB)$ and $l=\sqrt{\hbar c/|eB_\perp|}$ 
is the magnetic length. In principle, they could also depend on the quasiparticle energy $\omega$. 
Taking into account that $(\gamma^0)^2=\mathbb{I}$ and $(is_{\perp}\gamma^1\gamma^2)^2=\mathbb{I}$, 
the Dirac structure of the operator-valued functions $\hat{F}^+$ and $\hat\Sigma^+$ can be written in 
the following form:
\begin{eqnarray}
\hat{F}^{+} & = & f  + \gamma^0 g + i s_{\perp}\gamma^1\gamma^2 \tilde{g} + i s_{\perp}\gamma^0\gamma^1\gamma^2 \tilde{f} , 
\label{F-general}\\ 
\hat{\Sigma}^{+} & = & \tilde{\Delta} + \gamma^0 \mu + i s_{\perp}\gamma^1\gamma^2 \tilde{\mu} + i s_{\perp}\gamma^0\gamma^1\gamma^2 \Delta,
\label{Sigma-general}
\end{eqnarray}
where $f$, $\tilde{f}$, $g$, $\tilde{g}$, $\tilde{\Delta}$, $\Delta$, $\mu$, and $\tilde\mu$ are functions 
of only one operator, $(\bm{\pi}\cdot\bm{\gamma})^2l^2$. To a large degree the physical meaning of 
the corresponding operators should be clear from their Dirac structure and symmetry properties. 
In particular, the eigenvalues of the first four of them ($f$, $\tilde{f}$, $g$, and $\tilde{g}$) will play 
the role of generalized wave-function renormalizations, while the others will be playing the role of 
the generalized the MC and QHF order parameters, i.e., the Dirac (parity-even) and Haldane 
(parity-odd) masses (i.e., $\Delta$ and $\tilde{\Delta}$) and chemical potentials (i.e., $\mu$ and $\tilde\mu$). 
As we will see, such an interpretation is also supported by the role of the corresponding parameters
in the dispersion relations of quasiparticles, see Eqs.~(\ref{LLL-energy}) and (\ref{LL-energies}) below.

The three mutually commuting operators, $\gamma^0$, $is_{\perp}\gamma^1\gamma^2$, and 
$(\bm{\pi}\cdot\bm{\gamma})^2l^2$, allow a common basis of eigenstates, $|s_0, s_{12}, n\rangle$. 
The corresponding eigenvalues are $s_0=\pm1$, $s_{12}=\pm1$, and $-2n=-(2N+1+s_\perp s_{12})$, 
respectively, where $N=0,1,2,...$ is the quantum number of the orbital angular momentum and $s_{12}$ 
is the sign of the pseudospin projection. By making use of the complete set of eigenstates 
$|s_0, s_{12}, n\rangle$, one can derive a very convenient generalized Landau level representation 
(GLLR) of the (inverse) quasiparticle propagator (\ref{fullProp}). (For details of the derivation, see 
Appendix~A in Ref.~\cite{gorbar:2012ps}.) The final form of the {\it inverse} full propagator is given by
\begin{equation}
 G^{-1} (\omega;\mathbf{r},\mathbf{r}^\prime) = e^{i\Phi(\mathbf{r},\mathbf{r}^\prime)}  \tilde{G}^{-1} (\omega;\mathbf{r}-\mathbf{r}^\prime) ,
 \label{G-inv}
\end{equation}
where $\Phi(\mathbf{r},\mathbf{r}^\prime)= - s_\perp \frac{(x+x^\prime)(y-y^\prime)}{2l^2}$ is the well-known Schwinger phase
in an external magnetic field in the Landau gauge $\mathbf{A} = (0,B x)$. The translation invariant part 
of the inverse GLLR propagator is given by
\begin{eqnarray}
\tilde{G}^{-1} (\omega;\mathbf{r}) &=& -i \frac{e^{-\xi/2}}{2\pi l^2} \sum_{n=0}^{\infty}  \sum_{s_0=\pm1}\sum_{\sigma=\pm1}
\Bigg\{
s_0\omega L_{n}(\xi) + ( s_0 \mu_{n,\sigma} + \tilde{\Delta}_{n,\sigma} ) \left[ \delta_{-\sigma}^{s_0} L_{n}(\xi) + \delta_{+\sigma}^{s_0} L_{n-1}(\xi) \right] \nonumber\\
&&+ i \frac{v_F^2}{l^2} (\bm{\gamma} \cdot \mathbf{r} ) (f_{n,\sigma}-s_0 g_{n,\sigma}) L_{n-1}^1(\xi) 
\Bigg\} {\cal{P}}_{s_0, s_0\sigma} \, ,
 \label{G-inv-transl}
\end{eqnarray}
where $\xi = \mathbf{r}^2/(2l^2)$ and $L^\alpha_n(\xi)$ are the Laguerre polynomials. 
(Here, by definition, $L^0_n(\xi)\equiv L_n(\xi)$ and $L^\alpha_{n<0}(\xi)=0$.) In Eq.~(\ref{G-inv-transl}), 
we also used the following set of eigenstate projectors in the Dirac space: 
$P_{s_0,s_{12}}=\frac{1}{4}(1+s_0\gamma^0)(1+s_{12}is_\perp\gamma^1\gamma^2)$, 
as well as  the following shorthand notations for the linear combinations of the order 
parameters: 
\begin{equation}
\mu_{n,\sigma}=\mu_n + \sigma \tilde{\mu}_n, \qquad \tilde{\Delta}_{n,\sigma}= \tilde{\Delta}_n + \sigma \Delta_n, \qquad
f_{n,\sigma}=f_n + \sigma \tilde{f}_n, \qquad g_{n,\sigma}= g_n + \sigma \tilde{g}_n,
\label{8pars}
\end{equation}
where $\sigma \equiv s_0 s_{12}$. Note that the parameters $f_{n}$, $\tilde{f}_{n}$, $g_{n}$, 
$\tilde{g}_{n}$, $\tilde{\Delta}_{n}$, $\Delta_{n}$, $\mu_{n}$, and $\tilde\mu_{n}$ are associated 
with the $n$th Landau level. They are obtained by calculating the eigenvalues of the 
corresponding operators, introduced on the right-hand side of Eqs.~(\ref{F-general}) 
and (\ref{Sigma-general}).  

Similarly, the full GLLR propagator takes the form \cite{gorbar:2012ps} 
\begin{equation}
G (\omega;\mathbf{r},\mathbf{r}^\prime) = e^{i\Phi(\mathbf{r},\mathbf{r}^\prime)}  \tilde{G} (\omega;\mathbf{r}-\mathbf{r}^\prime) ,
 \label{G-prop}
\end{equation}
where the Schwinger phase is exactly the same as in Eq.~(\ref{G-inv}) and the translation invariant part of the 
propagator reads
\begin{eqnarray}
\tilde{G} (\omega;\mathbf{r}) &=&  i \frac{e^{-\xi/2}}{2\pi l^2} \sum_{n=0}^{\infty}  \sum_{s_0=\pm1}\sum_{\sigma=\pm1}
\Bigg\{
\frac{  s_0(\omega+\mu_{n,\sigma}) + \tilde{\Delta}_{ n,\sigma}  } {(\omega+\mu_{n,\sigma})^2-E_{n,\sigma}^2} 
\left[ \delta_{-\sigma}^{s_0} L_{n}(\xi) + \delta_{+\sigma}^{s_0} L_{n-1}(\xi) \right] 
\nonumber\\
&&+ i \frac{v_F^2}{l^2} ( \bm{\gamma} \cdot \mathbf{r} ) \frac{ f_{n,\sigma}-s_0 g_{n,\sigma} } 
{(\omega+\mu_{n,\sigma})^2-E_{n,\sigma}^2} L_{n-1}^1(\xi) \Bigg\}  {\cal{P}}_{s_0, s_0\sigma}\, ,
 \label{G-prop-transl}
\end{eqnarray}
The explicit form of the Landau level energies $E_{n,\sigma}$ are given by
\begin{eqnarray}
E_{0,\sigma}&=&\sigma\tilde{\Delta}_{0,\sigma}=\Delta_0 + \sigma\tilde{\Delta}_0, \label{LLL-energy}\\
E_{n,\sigma}&=& \sqrt{ 2n \left( f_{n,\sigma}^2-g_{n,\sigma}^2 \right) v_F^2/l^2 + \tilde{\Delta}_{n,\sigma}^2},
\qquad n\geq 1 .
\label{LL-energies}
\end{eqnarray}
The corresponding quasiparticles energies are determined by the location of the poles in the 
propagator (\ref{G-prop-transl}), namely $\omega_{0,\sigma}=-\mu_{0,\sigma}+E_{0,\sigma}$ 
and $\omega_{n,\sigma}^\pm=-\mu_{n,\sigma} \pm E_{n,\sigma}$ for $n\ge 1$. 

Note that the expressions for the Landau level energies in Eqs.~(\ref{LLL-energy}) and (\ref{LL-energies})
shed additional light on the physical meaning of the various dynamical parameters, used in the ansatz 
of the full quasiparticle propagator. In particular, as we can see, the combination of the wave-function
renormalization parameters $\sqrt{f_{n,\sigma}^2-g_{n,\sigma}^2}$ determines the renormalization 
of the quasiparticle velocity. Also, we see that the absolute value of $\tilde{\Delta}_{n,\sigma}$ 
plays the role of a mass.

\subsection{Schwinger-Dyson (gap) equation with static screening}
\label{SDEscreening}

In order to derive the GLLR form of the Schwinger-Dyson (gap) equation, we will start from the standard  
coordinate form of the gap equation in the random-phase approximation (RPA) and assume that the 
interaction is provided by the long-range Coulomb force subject to static screening effects. Such a 
consideration will amend the analysis of Ref.~\cite{gorbar:2012ps}, where the corresponding gap 
equation was analyzed in the approximation without screening. Our goal here is to illuminate the 
qualitative and quantitative role played by screening. 

It is important to emphasize that we will use a mean-field method that ignores the 
fluctuations of order parameters. While this is a common approximation used in numerous theoretical 
studies, there is no rigorous justification of its validity. Formally, the corresponding fluctuations can destroy 
any long-range order in 2D when $T\neq 0$ and, thus, prevent any spontaneous breakdown of continuous 
symmetries. We argue semirigorously that this may not be the case for finite-size, high-quality graphene devices 
fabricated on substrates. The substrate could make certain aspects of dynamics in graphene 
effectively three-dimensional and, thus, tame the dangerous fluctuations (at least on the length
scales of typical devices) that would destroy the long-range order in an ideal 2D graphene. 
Admittedly, however, this issue requires a more careful investigation in the future studies.

After taking into account the condition of overall charge neutrality in graphene, we arrive at the 
following gap equation for the translation invariant part of the quasiparticle propagator \cite{gorbar:2012ps}:
\begin{equation}
\tilde{G}^{-1}(\omega;\mathbf{r})=\tilde{S}^{-1}(\omega;\mathbf{r})
+e^2\gamma^0 \tilde{G}(\omega;\mathbf{r})\gamma^0 D(\omega;\mathbf{r}) ,
\label{gap-equation}
\end{equation}
where $D(\omega; \mathbf{r})$ is the time-like component of the gauge (photon) field propagator that 
contains all the information about the interaction. As stated before, we will use an approximation 
with Coulomb interaction that includes the effects of static screening. In essence, this is 
the instantaneous approximation \cite{Gorbar:2002iw} that neglects the retardation of the 
interaction. One might argue that this is a reasonable approximation because charge carriers 
are much slower than the speed of light. (See, however, Refs.~\cite{Pyatkovskiy:2010xz,
Gorbar:2012jc} suggesting that dynamical screening could affect the dynamics quantitatively.) Therefore, 
we use an energy independent polarization function, i.e., $\Pi(\omega,\mathbf{k})\simeq\Pi(0,\mathbf{k})$, 
in order to model the effects of screening in the photon propagator. In momentum space, the 
latter reads 
\begin{equation}
D(\omega,\mathbf{k})\approx D(0,\mathbf{k})=\frac{i}{\epsilon_0\left[ k+\Pi(0,\mathbf{k}) \right]},
\end{equation}
where $\epsilon_0$ is the dielectric constant, associated with the substrate. When the value of 
$\epsilon_0$ is large, the underlying dynamics becomes weakly coupled and the approximation 
for the gap equation (\ref{gap-equation}) should become reliable. This is not always the 
case in graphene. However, we expect that this may be applicable in the case of the highest quality 
graphene devices, fabricated on a thin hexagonal boron nitride substrate \cite{Young2012}.

A convenient explicit form of the polarization function 
in the presence of an external magnetic field was calculated in Refs.~\cite{Pyatkovskiy:2010xz}. 
In the approximation that neglects the wave-function renormalization and the Dirac masses, it reads
\begin{equation}
\Pi(0,\mathbf{k}) = \frac{e^2N_f}{16\pi l^2T} \sum_{n=0}^\infty \sum_{\lambda=\pm} 
\frac{Q_{nn}^{\lambda\lambda}(y)}{\cosh^2(\frac{\mu-\lambda M_n}{2T})}
- \frac{e^2N_f}{4\pi l^2} \sum_{\substack{n,n^\prime=0\\ \lambda n\neq\lambda^\prime n^\prime}}^\infty 
\sum_{\lambda,\lambda^\prime=\pm} Q_{n n^\prime}^{\lambda\lambda^\prime}(y)
\frac{n_F(\lambda M_n)-n_F(\lambda^\prime M_{n^\prime})}{\lambda M_{n}-\lambda^\prime M_{n^\prime}} ,
\end{equation}
where $M_n=\sqrt{2nv_F^2/l^2}$, $T$ is the temperature, and the explicit form of function 
$Q_{n,n^\prime}^{\lambda,\lambda^\prime}(y)$ is given by \cite{Pyatkovskiy:2010xz}
\begin{equation}
Q_{n,n^\prime}^{\lambda,\lambda^\prime}(y)=e^{-y}y^{|n-n^\prime|}
\left(
\sqrt{ \frac{(1+\lambda\lambda^\prime\delta_{0n_>})n_<!}{n_>!} } L_{n<}^{|n-n^\prime|}(y) + \lambda\lambda^\prime (1-\delta_{0n_<}) 
\sqrt{ \frac{(n_<-1)!}{(n_>-1)!} L_{n_<-1}^{|n-n^\prime|}(y) } 
\right)^2 .
\end{equation}
Here, by definition, $y=\mathbf{k}^2l^2/2$, $n_>=$max$(n,n^\prime)$ and $n_<=$min$(n,n^\prime)$.
We note that neglecting the wave-function renormalization and masses in the calculation of the 
polarization function should provide a reasonable approximation to leading order in weak coupling. 
Moreover, this may work also at moderate coupling because the bulk of the polarization effects 
appear to be determined by the total number of filled Landau levels and not as much by the details 
of the quasiparticle dispersion relations.

For the derivation of the GLLR form of the gap equations, we refer the reader to Appendix B in 
Ref.~\cite{gorbar:2012ps}. The final set of equations reads
\begin{eqnarray}
\mu_{n,\sigma}-\mu_{\rm e} -\sigma\tilde{\Delta}_{n,\sigma}
&=& \frac{\alpha\epsilon_l}{2} \sum_{n^\prime=0}^{\infty} \kappa_{n^\prime,n}^{(0)} 
\Big\{ n_F(E_{n^\prime,\sigma}-\mu_{n^\prime,\sigma}) - n_F(E_{n^\prime,\sigma}+\mu_{n^\prime,\sigma}) \nonumber\\
&&- \frac{\sigma\tilde{\Delta}_{n^\prime,\sigma}}{E_{n^\prime,\sigma}} 
\left[ n_F(\mu_{n^\prime,\sigma}-E_{n^\prime,\sigma}) - n_F(E_{n^\prime,\sigma}+\mu_{n^\prime,\sigma}) \right] \Big\} ,
\label{gap-eq-1}
\end{eqnarray}
for $n\ge 0$, and
\begin{eqnarray}
\mu_{n,\sigma}-\mu_{\rm e} +\sigma\tilde{\Delta}_{n,\sigma}
&=& \frac{\alpha\epsilon_l}{2} \sum_{n^\prime=1}^{\infty} \kappa_{n^\prime-1,n-1}^{(0)} 
\Big\{ n_F(E_{n^\prime,\sigma}-\mu_{n^\prime,\sigma}) - n_F(E_{n^\prime,\sigma}+\mu_{n^\prime,\sigma})
\nonumber\\
&&+ \frac{\sigma\tilde{\Delta}_{n^\prime,\sigma}}{E_{n^\prime,\sigma}} 
\left[ n_F(\mu_{n^\prime,\sigma}-E_{n^\prime,\sigma}) - n_F(E_{n^\prime,\sigma}+\mu_{n^\prime,\sigma}) \right] \Big\} ,
\label{gap-eq-2}
\\
f_{n,\sigma}
&=& 1+\frac{\alpha\epsilon_l}{2} \sum_{n^\prime=1}^{\infty} \frac{\kappa_{n^\prime-1,n-1}^{(1)}}{n} \frac{f_{n^\prime,\sigma}}{E_{n^\prime,\sigma}} 
\left[ n_F(\mu_{n^\prime,\sigma}-E_{n^\prime,\sigma}) - n_F(E_{n^\prime,\sigma}+\mu_{n^\prime,\sigma}) \right] ,
\label{gap-eq-3}
\\
g_{n,\sigma}
&=& \frac{\alpha\epsilon_l}{2} \sum_{n^\prime=1}^{\infty} \frac{\kappa_{n^\prime-1,n-1}^{(1)}}{n} \frac{g_{n^\prime,\sigma}}{E_{n^\prime,\sigma}} 
\left[ n_F(\mu_{n^\prime,\sigma}-E_{n^\prime,\sigma}) - n_F(E_{n^\prime,\sigma}+\mu_{n^\prime,\sigma}) \right]  ,
\label{gap-eq-4}
\end{eqnarray}
for $n\ge 1$. In these equations, we introduced a dimensionless coupling constant, $\alpha
\equiv e^2/\epsilon_0 v_F \approx 2.2$, which is an analog of the fine structure 
constant for suspended graphene. 

The effect of screening in the above set of gap equations is implicit. It comes only through the 
modified values of the kernel coefficients \cite{gorbar:2012ps},
\begin{equation}
\kappa_{m,n}^{(\rho)}=\int_0^\infty \frac{dk}{2\pi} \frac{kl {\cal L}_{m,n}^{(\rho)}(kl)}{k+\Pi(0,k)}, \quad
\mbox{with} \quad \rho=0,1.
\label{kappa-12}
\end{equation}
It is straightforward to calculate numerically the static polarization function $\Pi(0,k)$ as a function of 
dimensionless variable $kl$. We checked that, for sufficiently small and sufficiently large values of $kl$, 
the numerical results approach the expected asymptotical behavior at $k\to 0$ and $k\to \infty$, 
respectively, derived in Eqs.~(19) and (22) in Ref.~\cite{Pyatkovskiy:2010xz}. Then, by making use
of the definition in Eq.~(\ref{kappa-12}), we obtain the numerical values of the kernel coefficients 
$\kappa_{m,n}^{(\rho)}$ (with $\rho=0,1$). The results for the first few Landau levels are presented 
in Tables~\ref{tab-kappa-0} and \ref{tab-kappa-1}, respectively. By comparing these with the kernel 
coefficients in Ref.~\cite{gorbar:2012ps}, we observe that the screening effects substantially (i.e., by 
about a factor of $2$) decrease the values of $\kappa_{m,n}^{(\rho)}$. As we will see in the next section,
this change strongly affects the magnitude of the order parameters (i.e., the symmetry breaking Dirac 
masses and chemical potentials) in the QH states with fractional filling of Landau levels.

%%%%%%%%%%%%%%%%%%%%%%%%%%%%%%%%%%%%%%%%
%%%%%%%%%%%%%%%%%%%%%%%%%%%%%%%%%%%%%%%%
%%%%%%%%%%%%%%%%%%%%%%%%%%%%%%%%%%%%%%%%
\begin{table}
{%\setlength{\extrarowheight}{4pt}
\begin{ruledtabular}
\caption{The kernel coefficients $\kappa_{m,n}^{(0)}$ with the effects of static screening included.} 
\begin{tabular}{clllllllllll}
$\kappa_{m,n}^{(0)}$ & $m=0$ & $m=1$ & $m=2$ & $m=3$ & $m=4$ & $m=5$ & $m=6$ & $m=7$ & $m=8$ & $m=9$ & $m=10$  \\
\hline 
    0  &   0.0744 & 0.0235 & 0.0164 & 0.0136 & 0.0120 & 0.0109 & 0.0101 & 0.0094 & 0.0088 & 0.0083 & 0.0079  \\
    1  &   0.0235 & 0.0602 & 0.0222 & 0.0156 & 0.0129 & 0.0114 & 0.0104 & 0.0096 & 0.0090 & 0.0085 & 0.0081  \\
    2  &   0.0164 & 0.0222 & 0.0531 & 0.0211 & 0.0149 & 0.0124 & 0.0109 & 0.0100 & 0.0093 & 0.0087 & 0.0083 \\
    3  &   0.0136 & 0.0156 & 0.0211 & 0.0486 & 0.0202 & 0.0144 & 0.0120 & 0.0106 & 0.0097 & 0.0090 & 0.0085 \\
    4  &   0.0120 & 0.0129 & 0.0149 & 0.0202 & 0.0453 & 0.0195 & 0.0140 & 0.0116 & 0.0103 & 0.0094 & 0.0088 \\
    5  &   0.0109 & 0.0114 & 0.0124 & 0.0144 & 0.0195 & 0.0428 & 0.0189 & 0.0137 & 0.0113 & 0.0100 & 0.0092 \\
    6  &   0.0101 & 0.0104 & 0.0109 & 0.0120 & 0.0140 & 0.0189 & 0.0407 & 0.0184 & 0.0133 & 0.0111 & 0.0098 \\
    7  &   0.0094 & 0.0096 & 0.0100 & 0.0106 & 0.0116 & 0.0137 & 0.0184 & 0.0390 & 0.0180 & 0.0131 & 0.0109 \\
    8  &   0.0088 & 0.0090 & 0.0093 & 0.0097 & 0.0103 & 0.0113 & 0.0133 & 0.0180 & 0.0375 & 0.0176 & 0.0128 \\
    9  &   0.0083 & 0.0085 & 0.0087 & 0.0090 & 0.0094 & 0.0100 & 0.0111 & 0.0131 & 0.0176 & 0.0362 & 0.0172 \\
   10  &   0.0079 & 0.0081 & 0.0083 & 0.0085 & 0.0088 & 0.0092 & 0.0098 & 0.0109 & 0.0128 & 0.0172 & 0.0351
\label{tab-kappa-0} 
\end{tabular}
\end{ruledtabular}}
\end{table}
%%%%%%%%%%%%%%%%%%%%%%%%%%%%%%%%%%%%%%%%
%%%%%%%%%%%%%%%%%%%%%%%%%%%%%%%%%%%%%%%%
%%%%%%%%%%%%%%%%%%%%%%%%%%%%%%%%%%%%%%%%

%%%%%%%%%%%%%%%%%%%%%%%%%%%%%%%%%%%%%%%%
%%%%%%%%%%%%%%%%%%%%%%%%%%%%%%%%%%%%%%%%
%%%%%%%%%%%%%%%%%%%%%%%%%%%%%%%%%%%%%%%%
\begin{table}
{%\setlength{\extrarowheight}{4pt}
\begin{ruledtabular}
\caption{The kernel coefficients $\kappa_{m,n}^{(1)}$ with the effects of static screening included.} 
\begin{tabular}{clllllllllll}
$\kappa_{m,n}^{(1)}$ & $m=0$ & $m=1$ & $m=2$ & $m=3$ & $m=4$ & $m=5$ & $m=6$ & $m=7$ & $m=8$ & $m=9$ & $m=10$  \\
\hline 
    0  &  0.0509 & 0.0142 & 0.0084 & 0.0064 & 0.0055 & 0.0051 & 0.0048 & 0.0046 & 0.0043 & 0.0041 & 0.0039 \\
    1  &  0.0142 & 0.0901 & 0.0283 & 0.0172 & 0.0131 & 0.0111 & 0.0099 & 0.0093 & 0.0088 & 0.0084 & 0.0081 \\
    2  &  0.0084 & 0.0283 & 0.1244 & 0.0419 & 0.0260 & 0.0198 & 0.0167 & 0.0148 & 0.0137 & 0.0129 & 0.0123 \\
    3  &  0.0064 & 0.0172 & 0.0419 & 0.1553 & 0.0550 & 0.0346 & 0.0264 & 0.0222 & 0.0197 & 0.0181 & 0.0169 \\
    4  &  0.0055 & 0.0131 & 0.0260 & 0.0550 & 0.1839 & 0.0677 & 0.0431 & 0.0330 & 0.0276 & 0.0244 & 0.0224 \\
    5  &  0.0051 & 0.0111 & 0.0198 & 0.0346 & 0.0677 & 0.2107 & 0.0801 & 0.0515 & 0.0395 & 0.0331 & 0.0292 \\
    6  &  0.0048 & 0.0099 & 0.0167 & 0.0264 & 0.0431 & 0.0801 & 0.2361 & 0.0922 & 0.0597 & 0.0459 & 0.0385 \\
    7  &  0.0046 & 0.0093 & 0.0148 & 0.0222 & 0.0330 & 0.0515 & 0.0922 & 0.2602 & 0.1039 & 0.0678 & 0.0523 \\
    8  &  0.0043 & 0.0088 & 0.0137 & 0.0197 & 0.0276 & 0.0395 & 0.0597 & 0.1039 & 0.2834 & 0.1154 & 0.0758 \\
    9  &  0.0041 & 0.0084 & 0.0129 & 0.0181 & 0.0244 & 0.0331 & 0.0459 & 0.0678 & 0.1154 & 0.3056 & 0.1266 \\
  10  &  0.0039 & 0.0081 & 0.0123 & 0.0169 & 0.0224 & 0.0292 & 0.0385 & 0.0523 & 0.0758 & 0.1266 & 0.3271 
   \label{tab-kappa-1} 
\end{tabular}
\end{ruledtabular}}
\end{table}
%%%%%%%%%%%%%%%%%%%%%%%%%%%%%%%%%%%%%%%%
%%%%%%%%%%%%%%%%%%%%%%%%%%%%%%%%%%%%%%%%
%%%%%%%%%%%%%%%%%%%%%%%%%%%%%%%%%%%%%%%%

\section{Numerical Results}
\label{sec:numerical}

In this section, we present our numerical solutions to the GLLR gap equations (\ref{gap-eq-1}) through 
(\ref{gap-eq-4}) in the model with static screening effects. In the calculation, we use the Newtonian 
iteration algorithm and replace the original infinite set of gap equations with a truncated set of 
$n_{\rm max}=50$ equations. In accordance with such a truncation, we also impose a sharp 
cutoff in the summation over the Landau level index at $n_{\rm max}=50$. (The numerical tests 
reveal that the cutoff at $n_{\rm max}=100$ does not affect any qualitative features of the solutions
and only slightly changes the numerical results for the dynamical parameters in the first few 
Landau levels.)

Taking into account that each Landau level ($n\leq n_{\rm max}$) has two possible spin states 
and each of them is characterized by eight different dynamical parameters, see Eq.~(\ref{8pars}), 
we have a total of $16 n_{\rm max}$ independent parameters that should be determined by 
solving $16 n_{\rm max}$ coupled algebraic equations, see Eq.~(\ref{gap-eq-1}) through 
(\ref{gap-eq-4}). [Strictly speaking the number of independent parameters is $16 n_{\rm max}+4$,
where the additional four corresponds to the $n=0$ Landau level, which is special.] 

Before proceeding with the numerical analysis, it is useful to note that the coupled sets of  
gap equations for the two spin states, $s=\downarrow,\uparrow$, have exactly the same form. 
They differ only in the value of the chemical potential, i.e., $\mu_{\uparrow}=\mu_{\rm e} -Z$ and 
$\mu_{\downarrow}=\mu_{\rm e} +Z$, where $Z\equiv \mu_B B$ is the Zeeman energy. We also 
note that the gap equations are the same for $\sigma=\pm1$ when the dynamical parameters
$f_{n,\sigma}$, $g_{n,\sigma}$, $\mu_{n,\sigma}$ and $\sigma\tilde{\Delta}_{n,\sigma}$ are
treated as independent variables. These two observations allow us to greatly reduce the 
numerical cost of calculations. We first consider the generic set of equations in a finite 
range of chemical potentials and tabulate possible solutions for the above-mentioned 
four independent parameters. Usually, there exist multiple solutions that differ in their 
symmetry properties. The final solutions for the complete set of dynamical parameters 
with fixed values of $s=\downarrow,\uparrow$ and $\sigma=\pm1$ are obtained by combining 
the distinct generic solutions for properly shifted values of the chemical potential. 

In the numerical analysis, it is convenient to express all physical quantities in units of the 
Landau energy scale,
\begin{equation}
\epsilon_l = \sqrt{\hbar v_F^2|eB_\perp|/c} \approx 26 \sqrt{ B_\perp\, [\mbox{T}]}\, \mbox{meV}
\approx 300 \sqrt{ B_\perp\, [\mbox{T}]}\, \mbox{K},
\label{LandauScale}
\end{equation}
where the value of the magnetic field is measured in teslas. It may be appropriate to emphasize 
here that, while $\epsilon_l$ is determined by the component of the magnetic field orthogonal to 
the plane of graphene $\mathbf{B}_\perp$, the Zeeman energy $Z = \mu_B B\approx 5.8\times 10^{-2} 
B\, [\mbox{T}]\, \mbox{meV}\approx 0.67 B\, [\mbox{T}]\, \mbox{K}$ is proportional to the absolute 
value of $\mathbf{B}$. Despite the fact that $B_\perp \leq B$, the Zeeman energy is generically 
much smaller than the Landau energy scale (\ref{LandauScale}). This changes only in the case 
when the magnetic field becomes nearly parallel to the plane of graphene, i.e., when $B_\perp\ll B$. 
In the analysis below, we do not consider such a limiting case. In general, the dimensionless Zeeman 
energy is given by $z\equiv Z/\epsilon_l  = 2.2 \times 10^{-3}\sqrt{ B\, [\mbox{T}]/\cos\theta_{B}}$, 
where $\theta_{B}$ is the angle of the magnetic field tilt. In the numerical analysis below, we will
fix the value of the dimensionless Zeeman energy to be $z=0.015$. This formally corresponds to 
$B/\cos\theta_{B} \approx 46.5\,\mbox{T}$.

\subsection{Fermi velocity renormalization in weak magnetic field}

In a weak magnetic field, the effect of symmetry breaking dynamical parameters is expected 
to be negligible. However, even in this regime it is interesting to explore the implications of the 
long-range Coulomb interaction on the quantum Hall effect in graphene. In particular, the interaction is 
expected to renormalize the Fermi velocity, which can be extracted from the dynamically modified 
expressions for the Landau level energies \cite{gorbar:2012ps}. In absence of QHF and MC order 
parameters, as we see from Eq.~(\ref{LL-energies}), the energies are given by $E_n 
= f_{n} v_F\sqrt{2n} \epsilon_l$, implying that the renormalized Fermi velocity is determined 
by the wave-function renormalization parameters $f_{n}$, namely $v_{n,F}^{\rm (ren)} 
\equiv  f_{n} v_F$. Without screening of the Coulomb interaction, the numerical values 
for the wave-function renormalization parameters $f_{n}$ were previously reported in 
Ref.~\cite{gorbar:2012ps}. The corresponding results with the effects of screening 
were reported in Ref.~\cite{Miransky:2015ava} for the QH states with filling factors 
$\nu=\pm 4(k+1/2)$. In the latter study, in fact, the dynamical parameters $\mu_n$ and 
$\Delta_n$ were also properly accounted for. By noting that the values of $f_{n}$ 
depend not only on the chemical potential, but also on the Landau level index $n$, 
we conclude that the same is also true for the renormalized Fermi velocity. This is 
a very interesting theoretical prediction that could be easily tested in optical experiments, 
for example, via a systematic study of absorption/transmission lines for each of the 
QH states \cite{sadowski:2006prl,jiang:2007prl,orlita:2010sst}.

%%%%%%%%%%%%%%%%%%%%%%%%%%%%%%%%%%%%%%%%
%%%%%%%%%%%%%%%%%%%%%%%%%%%%%%%%%%%%%%%%
%%%%%%%%%%%%%%%%%%%%%%%%%%%%%%%%%%%%%%%%
\begin{table}
{%\setlength{\extrarowheight}{4pt}
\begin{ruledtabular}
\caption{Values of the wave-function renormalization $f_n$ for several values of the chemical potentials with 
screening effects considered, compared with the solutions in Ref.~\cite{gorbar:2012ps} with screening effects 
neglected in the parentheses. (Note that the corresponding data in Ref.~\cite{gorbar:2012ps} had typos 
that were corrected here.)} 
\begin{tabular}{ccccccc}
 & $f_1$ & $f_2$ & $f_3$ & $f_4$ & $f_5$ & $f_6$  \\
\hline   
  $|\mu_{\rm e} |<\sqrt{2}\epsilon_l$                                    &        1.084 (1.270)    &   1.072 (1.243)  &    1.065 (1.227)  &    1.060 (1.214)   &   1.057 (1.205)  &    1.054 (1.197)   \\
  $\sqrt{2}\epsilon_l<|\mu_{\rm e} |<\sqrt{4}\epsilon_l$       &        1.045 (1.194)    &    1.066 (1.224) &    1.063 (1.217)  &   1.059 (1.208)    &   1.056 (1.201)  &    1.054 (1.194)   \\
  $\sqrt{4}\epsilon_l<|\mu_{\rm e} |<\sqrt{6}\epsilon_l$       &        1.037 (1.166)    &    1.042 (1.177)   &  1.058 (1.200)  &   1.057 (1.199)   &   1.055 (1.194)   &    1.052 (1.189)   \\
  $\sqrt{6}\epsilon_l<|\mu_{\rm e} |<\sqrt{8}\epsilon_l$       &        1.033 (1.150)  &   1.035 (1.156)  &    1.039 (1.165)  &     1.052 (1.184)   &   1.052 (1.185)   &    1.051 (1.182)   \\
  $\sqrt{8}\epsilon_l<|\mu_{\rm e} |<\sqrt{10}\epsilon_l$     &        1.031 (1.138)  &   1.032 (1.142)   &   1.034 (1.148)   &    1.037 (1.156)   &  1.048 (1.172)   &     1.049 (1.174)   \\
  $\sqrt{10}\epsilon_l<|\mu_{\rm e} |<\sqrt{12}\epsilon_l$   &        1.029 (1.128)  &   1.030 (1.132)   &   1.031 (1.136)  &    1.032 (1.141)  &   1.035 (1.148)  &     1.045 (1.162)    \\
  $\sqrt{12}\epsilon_l<|\mu_{\rm e} |<\sqrt{14}\epsilon_l$   &       1.027 (1.121)  &   1.028 (1.123)   &    1.028 (1.127)   &   1.029 (1.130)   &   1.031 (1.135)  &    1.034 (1.141)  
\label{Table-fn}
\end{tabular}
\end{ruledtabular}}
\end{table}
%%%%%%%%%%%%%%%%%%%%%%%%%%%%%%%%%%%%%%%%
%%%%%%%%%%%%%%%%%%%%%%%%%%%%%%%%%%%%%%%%
%%%%%%%%%%%%%%%%%%%%%%%%%%%%%%%%%%%%%%%%

In the context of the optical transitions, the effect of interactions can be conveniently quantified 
by measuring the deviations of the measured energies of transitions from the free theory predictions
\cite{jiang:2007prl},
\begin{equation}
\Delta E_{n,n^\prime}=E_{n^\prime}\pm E_{n}=(\sqrt{2n^\prime}\pm\sqrt{2n})\epsilon_l 
+ \alpha\epsilon_l C_{n,n^\prime},
\end{equation}
where the minus sign corresponds to transitions between states with negative energies and states 
with positive energies. Strictly speaking, the above definition of $\Delta E_{n,n^\prime}$ assumes
transitions between states with the same spin. In the case of transitions between different spin states, 
an extra $\pm 2Z$ correction should be added on the right-hand side. Here, the information about 
the wave-function renormalization 
is captured by the following set of dimensionless parameters $C_{n,n^\prime}$, i.e., 
\begin{equation}
C_{n,n^\prime}=\frac{\sqrt{2n^\prime}}{\alpha}(f_{n^\prime}-1)\pm \frac{\sqrt{2n}}{\alpha}(f_n-1).
\label{Cnn}
\end{equation}
As we see, in absence of symmetry breaking, which is the case in weak magnetic fields, 
the values of parameters $C_{n,n^\prime}$ are directly related to the quasiparticle velocity 
renormalizations. As we claim here,  for each QH state, the corresponding renormalizations 
are functions of the  Landau level index $n$. In experiment, the complete set of parameters 
$f_n$ with $n\geq 1$ could be extracted by measuring the transition energies between the 
lowest Landau level (which is free from from the corresponding renormalization effect) and 
higher Landau levels. The values of $f_n$, extracted in this way, would be sufficient to 
calculate the values of $C_{n,n^\prime}$ for transitions between various higher Landau levels.
The latter could be also compared to the actual measurements and, in the case of agreement, 
one would have a nontrivial test of the self-consistency of the GLLR ansatz used here. From 
a theoretical point of view, it will be perhaps even more interesting if deviations from the 
relations in Eq.~(\ref{Cnn}) are observed.

In a general case with nonvanishing QHF and MC order parameters, the expressions for 
$C_{n,n^\prime}$ parameters should be corrected because the Landau level energies are modified, see
Eqs.~(\ref{LLL-energy}) and (\ref{LL-energies}). The magnitude of the corresponding corrections 
is of the order of $|\mu_{n,\sigma}-\mu_{\rm e} |$ and $\tilde{\Delta}_{n,\sigma}^2/(\sqrt{n}\epsilon_l)$ 
for Landau levels $n\geq 1$. (Note that the correction due to the QHF order parameter should 
be interpret as part of the energy measured with respect to the thermodynamical potential 
$\mu_{\rm e} $.)  In the case of transitions to/from the lowest Landau level ($n=0$), 
the corrections are $\pm \sigma \tilde{\Delta}_{0,\sigma}$, see Eq.~(\ref{LLL-energy}). Below, 
we present our numerical results for the parameters $C_{n,n^\prime}$ in several QH states 
in both approximations, i.e., with and without inclusion of the QHF and MC order parameters. 

By utilizing the same value of the cutoff, $n_{\rm max}=100$, in the numerical calculation as in Ref.~\cite{gorbar:2012ps}, 
but also including the effects of screening, we straightforwardly obtain the wave-function renormalization parameters 
$f_n$. The corresponding zero temperature results for the first few Landau levels are presented in Table~\ref{Table-fn}. 
We note that they differ slightly from those in Ref.~\cite{Miransky:2015ava} because here all dynamical parameters 
such as $\mu_n$ and $\Delta_n$ were neglected. For comparison, in Table~\ref{Table-fn} we also list in the 
parentheses the numerical results from Ref.~\cite{gorbar:2012ps}, obtained without the screening effects. 
Different rows in Table~\ref{Table-fn} correspond to different choices of the chemical potential in the gaps 
between a completely filled $n$th Landau level and a completely empty $(n+1)$th Landau level. 

By comparing the results with and without screening in Table~\ref{Table-fn}, we find that the effect of 
wave function renormalization goes from about $14\%$ to $27\%$ down to about $3\%$ to $8\%$, which 
appears to be even smaller than the prediction in Ref.~\cite{lyengar:2007prb}. It is curious to explore in 
detail if the logarithmic running of the coupling constant could explain such a difference. By making use of 
the definition in Eq.~(\ref{Cnn}) and our numerical results for various QH states, we readily calculate 
$C_{n,n^\prime}$ parameters. 

The representative results for $C_{n,n^\prime}$ parameters are presented in Tables~\ref{Table-Cnn-nu=2} 
and \ref{Table-Cnn-nu=6} for the QH states with the filling factors $\nu=2$ and $\nu=6$. Because of the 
particle-hole symmetry, the results for the corresponding negative filling factors can be obtained as follows:
$C_{n,n^\prime}(-\nu) = C_{-n^\prime, -n}(\nu)$. These results appear to be about 
two to four times smaller than the results in the absence of screening \cite{gorbar:2012ps}. In calculation, 
we took into account the effect of nonvanishing QHF and MC order parameters. For comparison, in 
parenthesis we also show the results for the same parameters in the approximation with the QHF 
and MC order parameters neglected. It appears that the role of such parameters is not negligible.

\begin{table}
\begin{ruledtabular}
\caption{Parameters $C_{n,n^\prime}$ in the QH state with the filling factor $\nu=2$ in the model with 
static screening. The values in parentheses are obtained in an approximation with the wave-function 
renormalization effects included, but all QHF and MC order parameters neglected.} 
\begin{tabular}{ccccc}
$C_{n,n^\prime}$ & $n^\prime=1$ & $n^\prime=2$ & $n^\prime=3$ & $n^\prime=4$ \\
\hline 
   $n=0$   &    0.082 (0.054)  &  0.093 (0.065)  &  0.099 (0.072)  &  0.103 (0.077)  \\
   $n=-1$  &    0.100 (0.108)  &  0.111 (0.119)  &  0.117 (0.126)  &  0.122 (0.131)  \\
   $n=-2$  &    0.108 (0.119)  &  0.119 (0.130)  &  0.125 (0.138)  &  0.129 (0.143)  \\
   $n=-3$  &    0.113 (0.126)  &  0.124 (0.138)  &  0.130 (0.144)  &  0.134 (0.150)  \\
   $n=-4$  &    0.116 (0.131)  &  0.127 (0.143)  &  0.133 (0.150)  &  0.138 (0.220)  \\
\end{tabular}
\label{Table-Cnn-nu=2}
\end{ruledtabular}
\end{table}

\begin{table}
\begin{ruledtabular}
\caption{Parameters $C_{n,n^\prime}$ in the QH state with the filling factor $\nu=6$ in the model with 
static screening. The values in parentheses are obtained in an approximation with the wave-function 
renormalization effects included, but all QHF and MC order parameters neglected.} 
\begin{tabular}{cccc}
$C_{n,n^\prime}$ & $n^\prime=2$ & $n^\prime=3$ & $n^\prime=4$ \\
\hline 
   $n=1$     &  0.054 (0.031)  &  0.066 (0.041)  &  0.073 (0.047)  \\   
   $n=0$     &  0.088 (0.060) &  0.100 (0.070)  &  0.107 (0.076)  \\
   $n=-1$    &  0.103 (0.089)  &  0.115 (0.100)  &  0.122 (0.105)  \\
   $n=-2$    &  0.109 (0.121)  &  0.121 (0.131)  &  0.128 (0.136)  \\
   $n=-3$    &  0.113 (0.131)  &  0.125 (0.141)  &  0.132 (0.147)  \\
   $n=-4$    &  0.116 (0.136)  &  0.128 (0.147)  &  0.135 (0.152)  \\
\end{tabular}
\label{Table-Cnn-nu=6}
\end{ruledtabular}
\end{table}

\subsection{Quantum Hall states in strong magnetic field}

In this section, we study QH states with different integer filling factors at zero temperature. We will 
start by first considering the states associated with the $n=0$ and $n=1$ Landau levels. We will show, 
in particular, that the qualitative features of the phase diagram obtained in Ref.~\cite{gorbar:2012ps} 
remain qualitatively the same after the inclusion of the static screening. At the same time, the 
quantitative changes will be substantial. 

As stated earlier, the gap equations (\ref{gap-eq-1}) through (\ref{gap-eq-4}) allow a large number 
of solutions with different types of Haldane/Dirac masses ($\Delta_n$, $\tilde{\Delta}_n$) and chemical 
potentials ($\mu_n$, $\tilde\mu_n$). In order to identify the true ground state among them, we compare
their free energies. (For the explicit expression of the free energy, see Appendix C in Ref.~\cite{gorbar:2012ps}.)
In the model at hand, the choice of the corresponding states is strongly affected by the Zeeman 
interaction energy, which is one of the main factors in driving the vacuum alignment. Taking into 
account that there may exist a large number of other symmetry breaking effects, e.g., various 
on-site repulsion interaction terms \cite{Aleiner:2007va}, the actual nature of the ground states 
should be accepted with caution. Nevertheless, the study below is quite informative: it reveals 
the quantitative effects of the static screening and role of the running of the dynamical parameters 
in the QH regime of graphene.  

From the symmetry viewpoint, there are two types of Dirac masses and chemical potentials for
quasiparticles of each spin orientation, $s=\downarrow,\uparrow$. The parameters of the first type 
(i.e., $\Delta_n$ and $\mu_n$) are singlets, while the parameters of the second type (i.e., $\tilde{\Delta}_n$ 
and $\tilde\mu_n$) are triplets with respect to the flavor $U_{s}(2)$ subgroups. It is natural 
to expect that the states with different symmetry properties compete. However, it should be 
emphasized that the corresponding competition is not necessarily between the MC and QHF 
scenarios because both types of order parameters may belong to the same representations of 
the flavor symmetry. In fact, as our results show, the two types of order parameters generically 
coexist in all QH states. (This was also emphasized in Ref.~\cite{gorbar:2012ps}.)

From general considerations based on the structure of the Landau levels, it is expected that there exist 
at least four different classes of QH states with the following series of filling factors: (i) $\nu=4n+2$, (ii) $\nu=4n$ 
(with a possible exclusion of the rather unique $\nu =0$ state in a class of its own), (iii) $\nu=4n-1$, 
and (iv) $\nu=4n+1$. 

The first series of states with the filling factors $\nu=4n+2$ describes the ``normal" QH states with the 
complete filling of the (nearly) degenerate Landau levels. They do not have or require symmetry breaking 
and are resolved even at relatively weak magnetic fields. The simplest realization of the $\nu=4n$ states 
could be provided by a dynamically enhanced Zeeman splitting of Landau levels. In this case, the ground 
states have $U_{\downarrow}(2)\times U_{\uparrow}(2)$ symmetry. While this is also the prediction of the 
model at hand, we should emphasize that other realizations of the $\nu=4n$ QH states are possible. In fact, 
the $\nu=0$ state, which formally belongs to this series, is likely to have a different origin \cite{Kharitonov}. 
The remaining two series of states with the filling factors $\nu = 4n \pm 1$ are less controversial. They 
require spontaneous symmetry breaking at least down to $U_{\downarrow}(1)\times U_{\uparrow}(2)$ 
or $U_{\downarrow}(2)\times U_{\uparrow}(1)$. 

In our analysis at sufficiently small values of the chemical potential, we find a number of different 
solutions, associated with the lowest Landau level ($n=0$) and integer filling factors $\nu =0, \pm1, \pm2$. 
The order of appearance and the competition of different types of solutions appear to be the same as 
in Ref.~\cite{gorbar:2012ps}, see Figure~3 there. By taking into account the screening of the Coulomb 
interaction, however, we find that the actual values of dynamical parameters change substantially.
The corresponding results are listed in Table~\ref{Table-n0}. For comparison, in parenthesis we also 
list the previous results in the model without screening \cite{gorbar:2012ps}. As we see, the effect of 
screening is to suppress the relevant dynamical parameters by about a factor of $3$. The same is 
true for the magnitude of the energy gaps in the states with the filling factors $\nu =0, \pm 1$.

%%%%%%%%%%%%%%%%%%%%%%%%%%%%%%%%%%%%%%%%
%%%%%%%%%%%%%%%%%%%%%%%%%%%%%%%%%%%%%%%%
%%%%%%%%%%%%%%%%%%%%%%%%%%%%%%%%%%%%%%%%
\begin{table}
{%\setlength{\extrarowheight}{4pt}
\begin{ruledtabular}
\caption{Gap parameters for the solutions when the Fermi energy close to the lowest Landau level. The solutions for $n_{\rm max}=50$ 
with screening effects considered are compared with the solutions in Ref.~\cite{gorbar:2012ps} for $n_{\rm max}=5$ 
neglecting screening effects in the parentheses.} 
\begin{tabular}{ccccc}
$\nu$ & $\tilde{\Delta}_{0,\uparrow}^{\rm eff}$ & $\tilde{\Delta}_{0,\downarrow}^{\rm eff}$ & $\Delta_{0,\uparrow}$ &  $\Delta_{0,\downarrow}$   \\
\hline   
       -2 &  0.000 (0.000)  & 0.000 (0.000)  &  0.082 (0.227) &  0.082 (0.227) \\
       -1 &  0.000 (0.000)  & 0.082 (0.227) &  0.082 (0.227) &  0.000 (0.000) \\
        0  & 0.000 (0.000) & 0.000 (0.000) &  0.082 (0.227) & -0.082 (-0.227) \\
        1  & 0.082 (0.227) & 0.000 (0.000) &  0.000 (0.000) & -0.082 (-0.227) \\
        2 &  0.000 (0.000)  & 0.000 (0.000) & -0.082 (-0.227) & -0.082 (-0.227) 
\label{Table-n0}
\end{tabular}
\end{ruledtabular}}
\end{table}
%%%%%%%%%%%%%%%%%%%%%%%%%%%%%%%%%%%%%%%%
%%%%%%%%%%%%%%%%%%%%%%%%%%%%%%%%%%%%%%%%
%%%%%%%%%%%%%%%%%%%%%%%%%%%%%%%%%%%%%%%%

The analysis of the QH states, associated with the $n=1$ Landau level, is done 
in the same way. Here again, the order of appearance and the competition of different 
types of solutions appear to be exactly the same as in Ref.~\cite{gorbar:2012ps}, see 
Figure~4 there. The actual values of the dynamical parameters in the model with 
screening are listed in Table~\ref{Table-n1}. The corresponding results in the model 
without screening are given in parenthesis. By comparing the two sets of data, we
find that screening leads to a suppression of the relevant dynamical parameters by 
a factor of $3$ to $5$. 

%%%%%%%%%%%%%%%%%%%%%%%%%%%%%%%%%%%%%%%%
%%%%%%%%%%%%%%%%%%%%%%%%%%%%%%%%%%%%%%%%
%%%%%%%%%%%%%%%%%%%%%%%%%%%%%%%%%%%%%%%%
\begin{table}
{%\setlength{\extrarowheight}{4pt}
\begin{ruledtabular}
\caption{Gap parameters for the solutions when the Fermi energy close to the Landau level at $n=1$. The solutions for $n_{\rm max}=50$ 
with screening effects considered are compared with the solutions in Ref.~\cite{gorbar:2012ps} for $n_{\rm max}=5$ 
neglecting screening effects in the parentheses.} 
\begin{tabular}{ cccccccccccc }
  &     spin    &    $\tilde{\Delta}_{0}^{\rm eff}$  &    $\Delta_{0}$  &  $f_1$  &  $\mu_1 - \mu_s$  &  $\Delta_1$  &  $\tilde{\Delta}_1$  &   $f_2$   &  $\mu_2 - \mu_s$  &  $\Delta_2$  &   $\tilde{\Delta}_2$   \\[1ex] \hline 
\multirow{2}{*}
{ $\nu=3$ } &  \multirow{2}{*}{$\uparrow$}    &       0.000    &    -0.082      &       1.078    &      0.013     &      -0.014   &       0.000     &      1.065     &      0.009     &    -0.010     &     0.000     \\
                     &                                                             &    (0.000)    &    (-0.227)   &     (1.143)    &    (0.053)   &    (-0.068)   &     (0.000)    &     (1.112)    &    (0.040)    &   (-0.052)   &     0.000     \\[1ex] %\hline
                     &  \multirow{2}{*}{$\downarrow$}         &       -0.013   &     -0.095     &      1.058      &    0.050     &      -0.010   &       0.004      &      1.062    &       0.022     &    -0.009     &     0.000     \\ 
                     &                                                            &   (-0.051)    &    (-0.278)    &     (1.105)    &    (0.148)   &   (-0.049)    &     (0.018)    &     (1.102)    &    (0.091)    &   (-0.046)    &    (0.006)    \\ [1ex] \hline
\multirow{2}{*}
{ $\nu=4$ } &  \multirow{2}{*}{$\uparrow$}    &        0.000   &     -0.082     &      1.078      &       0.013    &    -0.014     &      0.000      &     1.065     &      0.009    &     -0.010     &       0.000      \\
                     &                                                             &      (0.000)  &   ( -0.227)   &     (1.143)    &    (0.053)    &    (-0.068)   &     (0.000)    &    (1.112)    &    (0.040)   &    (-0.052)   &     (0.000)     \\[1ex] %\hline
                     &  \multirow{2}{*}{$\downarrow$}         &        0.000    &     -0.108    &       1.038     &     0.087      &    -0.006     &       0.000     &     1.059      &     0.034    &     -0.009     &       0.000     \\
                     &                                                             &      (0.000)  &    (-0.330)   &     (1.066)    &    (0.243)    &   (-0.031)    &      (0.000)   &    (1.093)    &    (0.143)   &    (-0.039)   &      (0.000)     \\  [1ex] \hline
\multirow{2}{*}
{ $\nu=5$ } &  \multirow{2}{*}{$\uparrow$}    &       -0.013    &     -0.095    &        1.058    &     0.049      &      -0.010   &      0.004     &     1.062       &     0.021     &    -0.009      &     0.000     \\
                     &                                                             &     (-0.051)  &     (-0.278)  &      (1.105)   &    (0.148)    &     (-0.049)  &    (0.018)    &    (1.102)     &    (0.091)    &   (-0.046)    &   (0.006)    \\[1ex] %\hline 
                     &  \multirow{2}{*}{$\downarrow$}         &         0.000   &    -0.108     &         1.038    &     0.087     &      -0.006    &     0.000      &     1.059      &    0.034       &   -0.009      &      0.000      \\
                     &                                                             &       (0.000)   &    (-0.278)    &    (1.066)    &    (0.245)   &    (-0.031)   &    (0.000)    &    (1.093)    &    (0.143)    &   (-0.039)     &    (0.000)       \\  [1ex] \hline
\multirow{2}{*}
{ $\nu=6$ } &  \multirow{2}{*}{$\uparrow$}    &           0.000    &      -0.108     &   1.038        &     0.087    &     -0.006     &       0.000     &     1.059    &       0.034    &     -0.009     &      0.000  \\
                     &                                                             &         (0.000)   &    (-0.330)    &    (1.066)    &    (0.243)   &   (-0.031)    &      (0.000)   &    (1.093)   &     (0.142)    &   (-0.039)   &     (0.000)    \\[1ex] 
                     &  \multirow{2}{*}{$\downarrow$}         &           0.000     &     -0.108      &  1.038        &      0.087    &     -0.006    &       0.000     &     1.059     &      0.034     &    -0.009     &      0.000  \\
                     &                                                             &         (0.000)   &    (-0.331)    &    (1.066)    &    (0.245)   &   (-0.031)    &      (0.000)   &    (1.093)   &     (0.143)    &   (-0.039)   &     (0.000)      
\label{Table-n1}
\end{tabular}
\end{ruledtabular}}
\end{table}
%%%%%%%%%%%%%%%%%%%%%%%%%%%%%%%%%%%%%%%%
%%%%%%%%%%%%%%%%%%%%%%%%%%%%%%%%%%%%%%%%
%%%%%%%%%%%%%%%%%%%%%%%%%%%%%%%%%%%%%%%%

Because of the long-range nature of the Coulomb interaction, the dynamical QHF 
($\mu_n$, $\tilde\mu_n$) and MC ($\Delta_n$, $\tilde{\Delta}_n$) order parameters
are nontrivial functions of the Landau level index $n$. The corresponding ``running"
of the dynamical parameters is an important feature of the model at hand. From 
theoretical viewpoint, it is essential to provide a realistic description of the low-energy 
dynamics in graphene, where the role of order parameters diminishes with increasing 
the quasiparticle energy. This is in contrast to a common mean-field analysis of models 
with pointlike interactions, where the order parameters affect either (i) only the nearest 
filled Landau level or (ii) all levels in the same way.   

\begin{figure}[ht]
\centering
\includegraphics[width=0.45\textwidth]{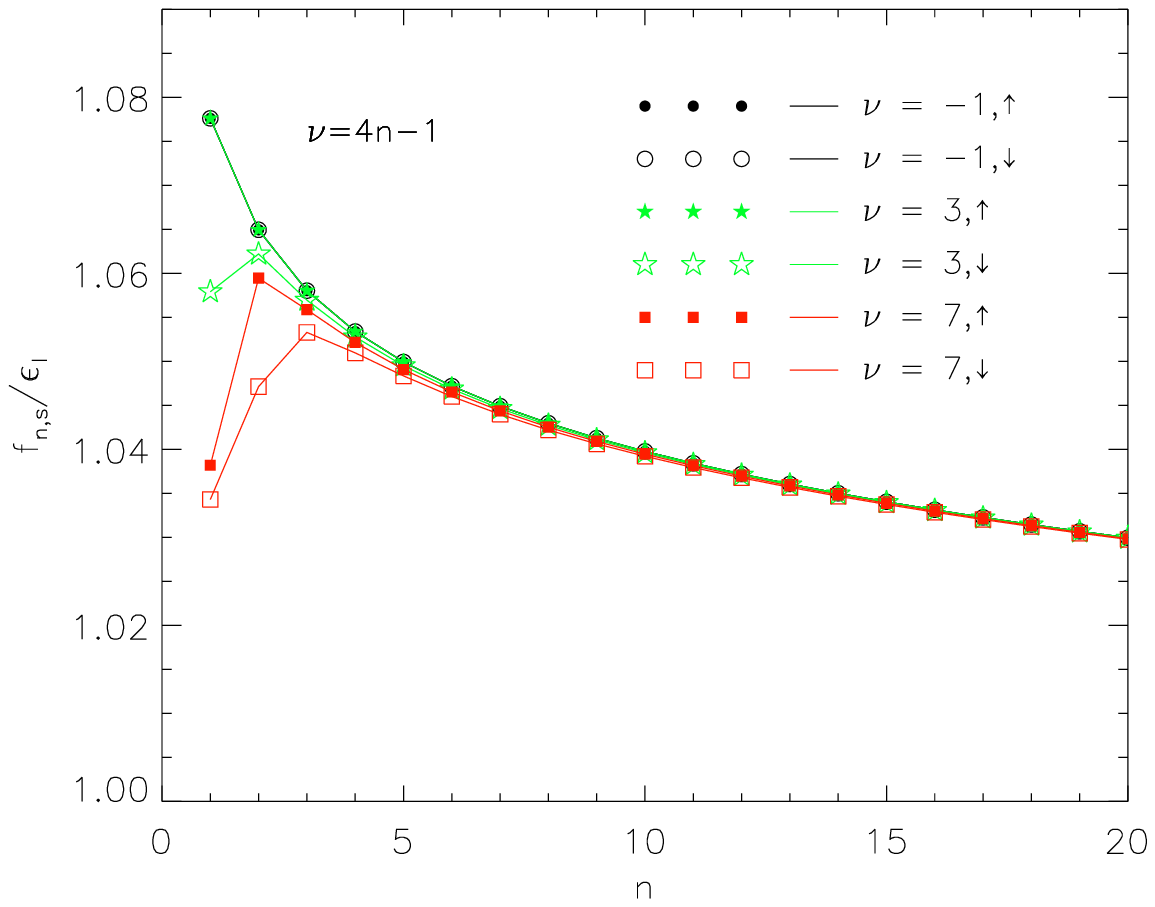}
\includegraphics[width=0.45\textwidth]{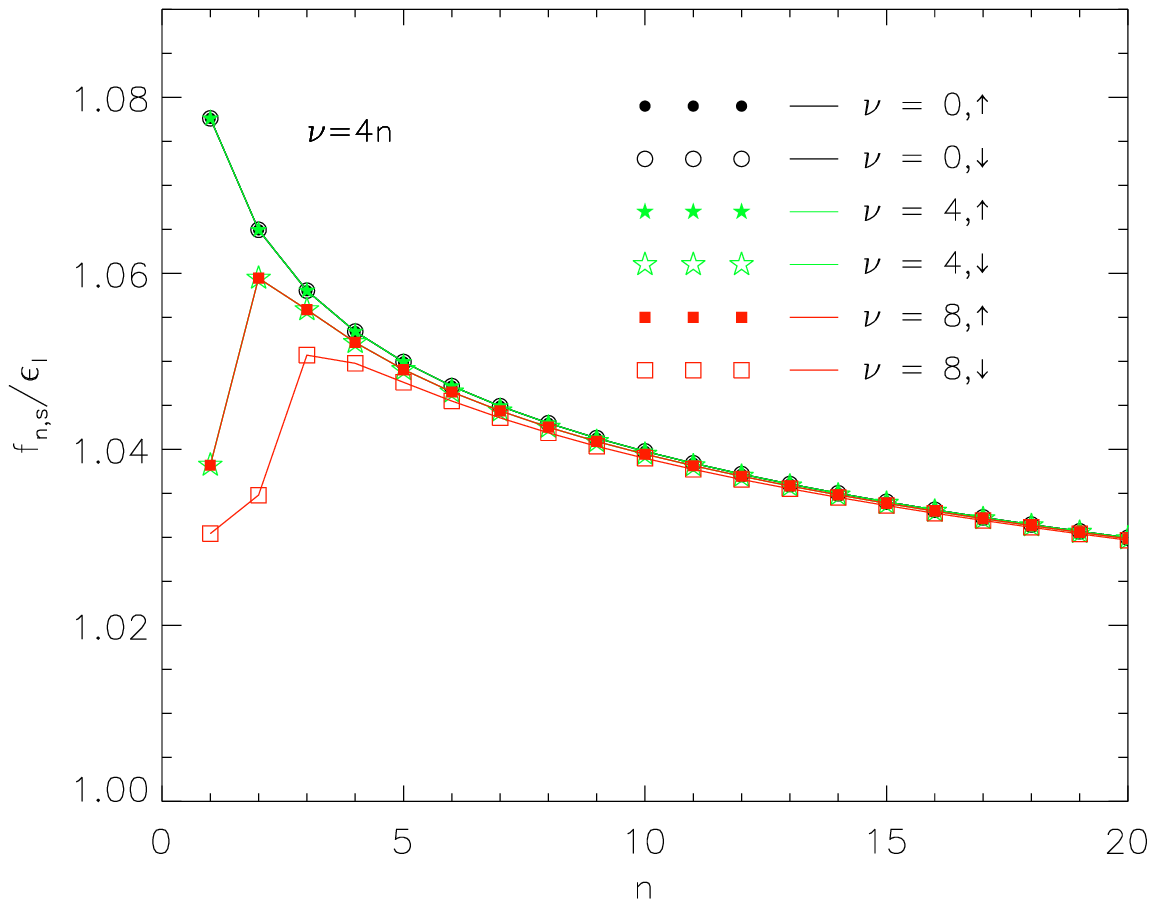}\\
\includegraphics[width=0.45\textwidth]{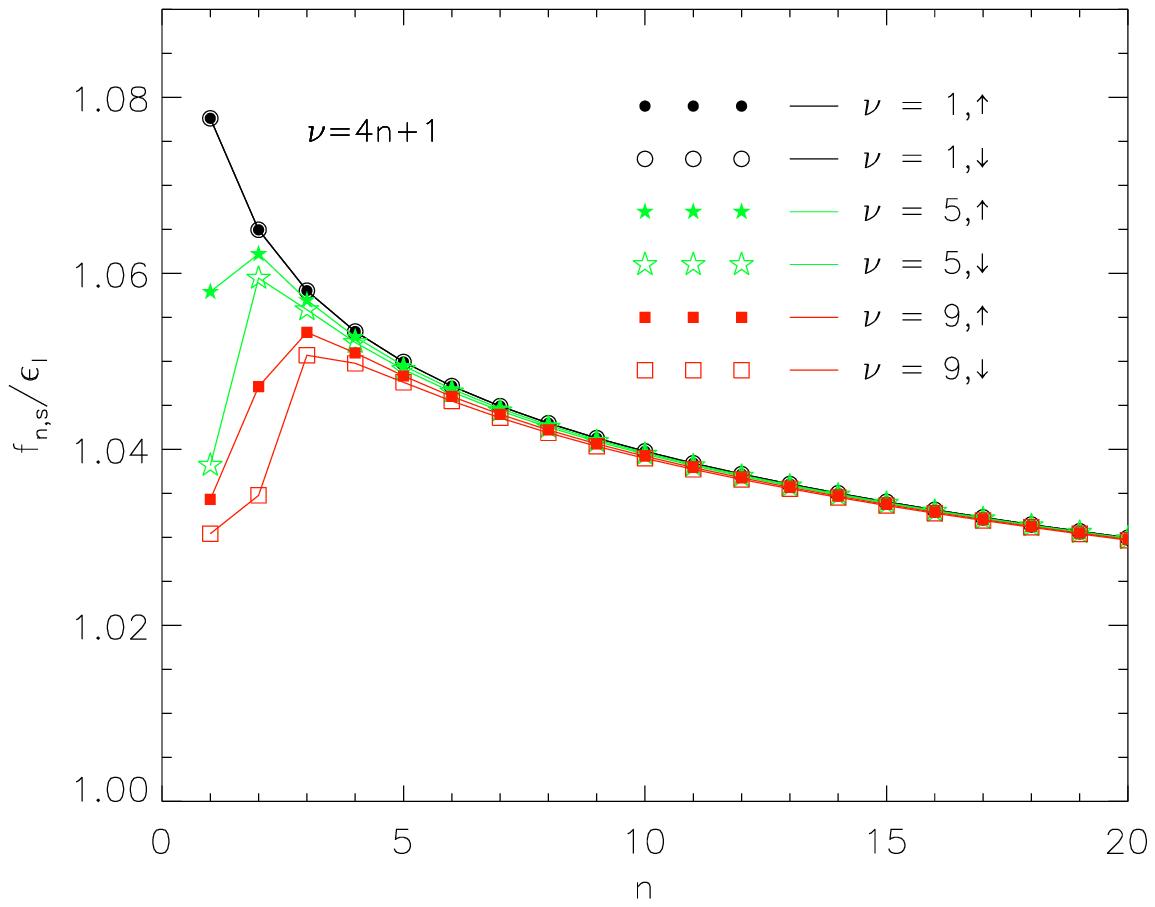}
\includegraphics[width=0.45\textwidth]{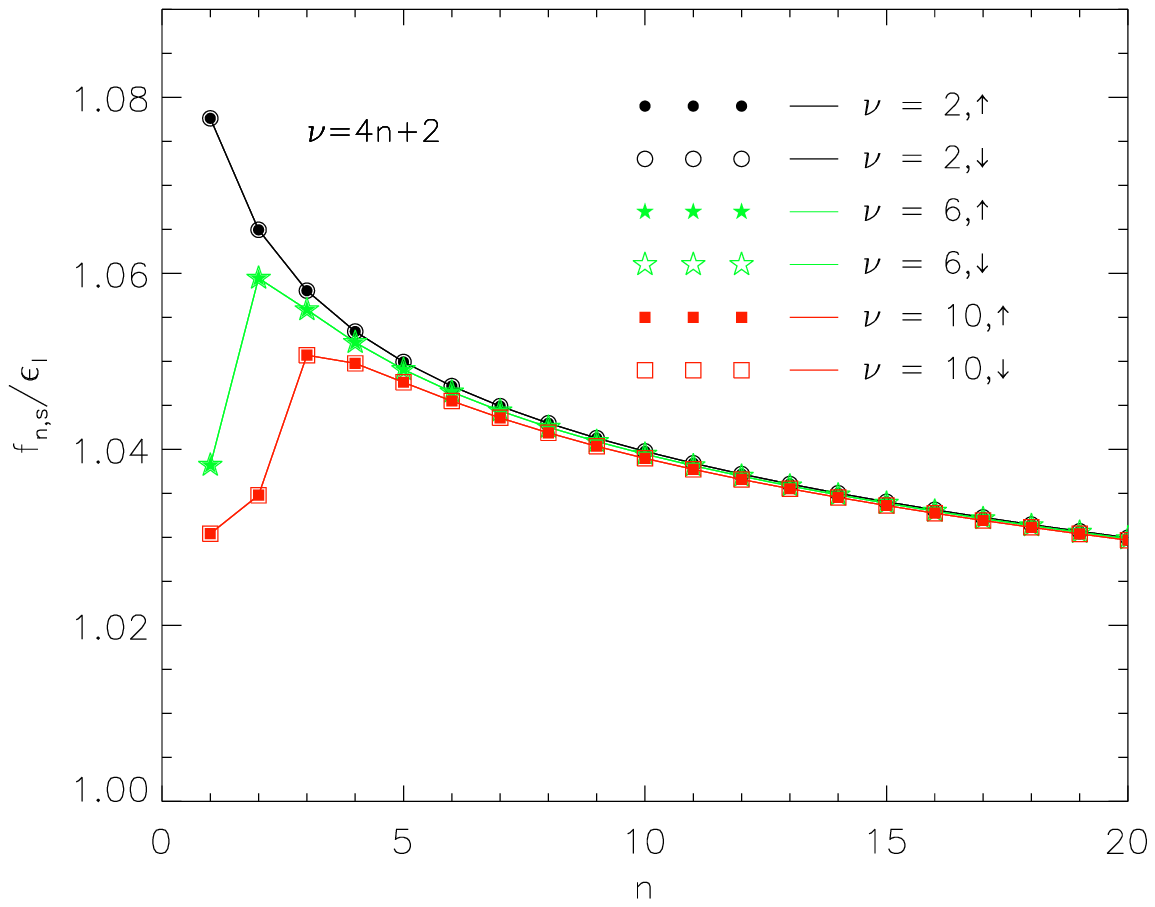}
\caption{(Color online) 
The wave-function renormalization parameters $f_{n,s}$ as functions of the Landau level index $n$ 
for four different types of QH states with filling factors $\nu=4n-1$,  $\nu=4n$,  $\nu=4n+1$, and 
$\nu=4n+2$.}
\label{fig_f-n}
\end{figure}

\begin{figure}[ht]
\centering
\includegraphics[width=0.45\textwidth]{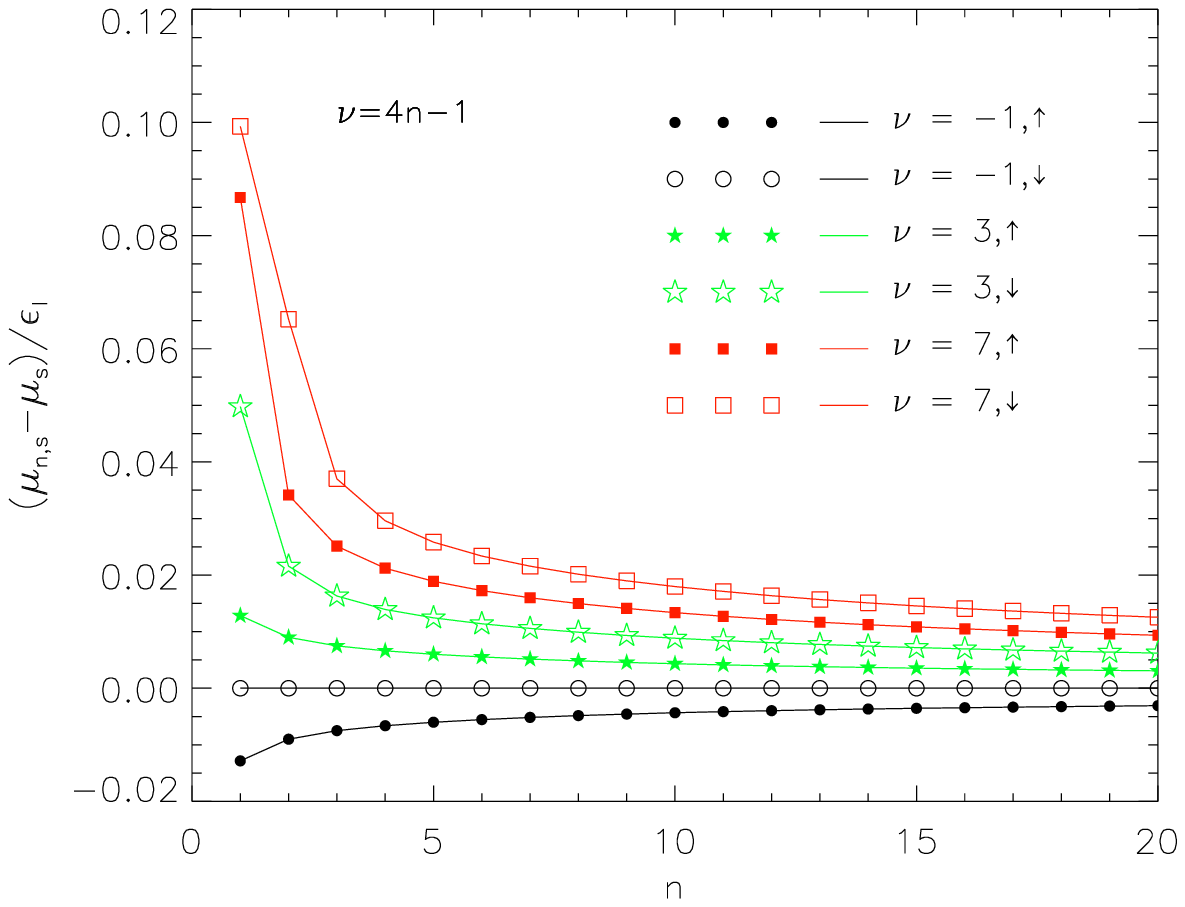}
\includegraphics[width=0.45\textwidth]{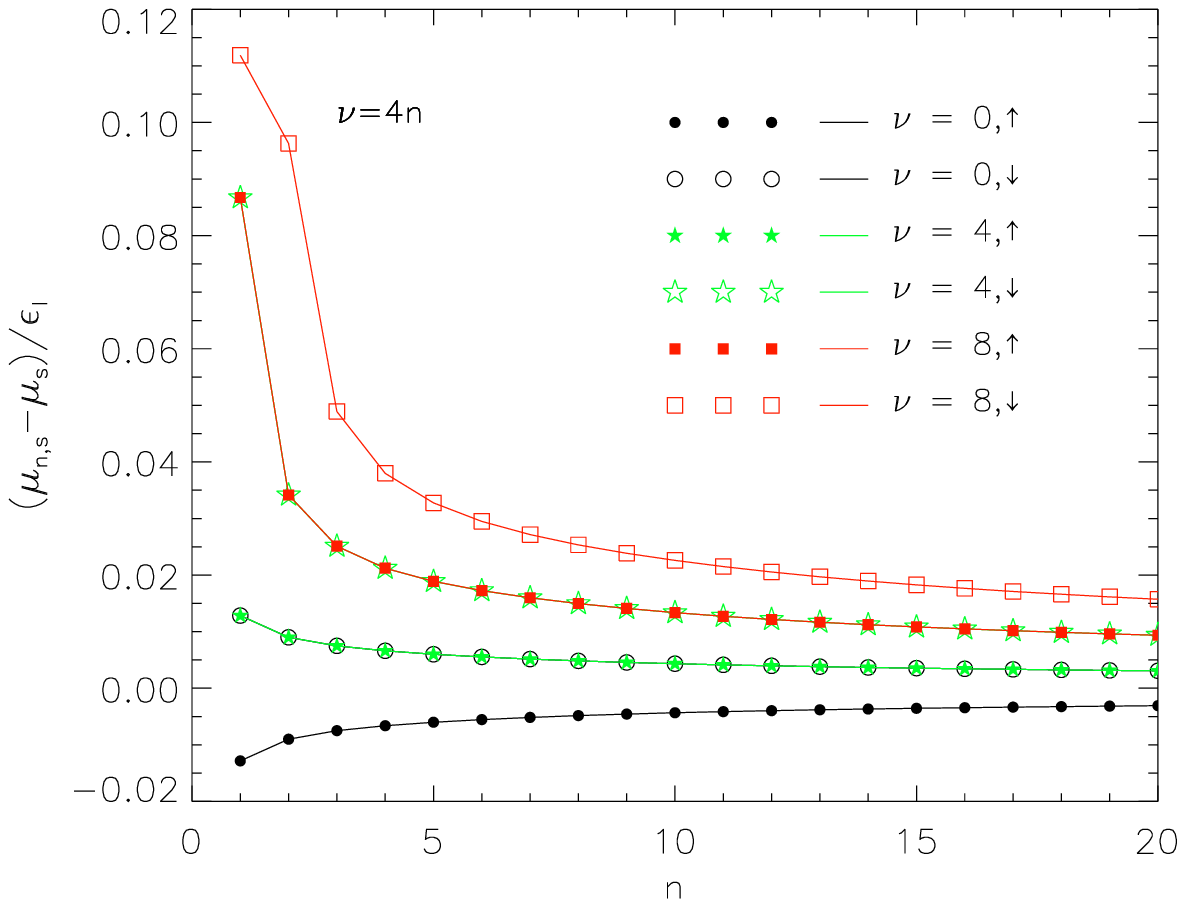}\\
\includegraphics[width=0.45\textwidth]{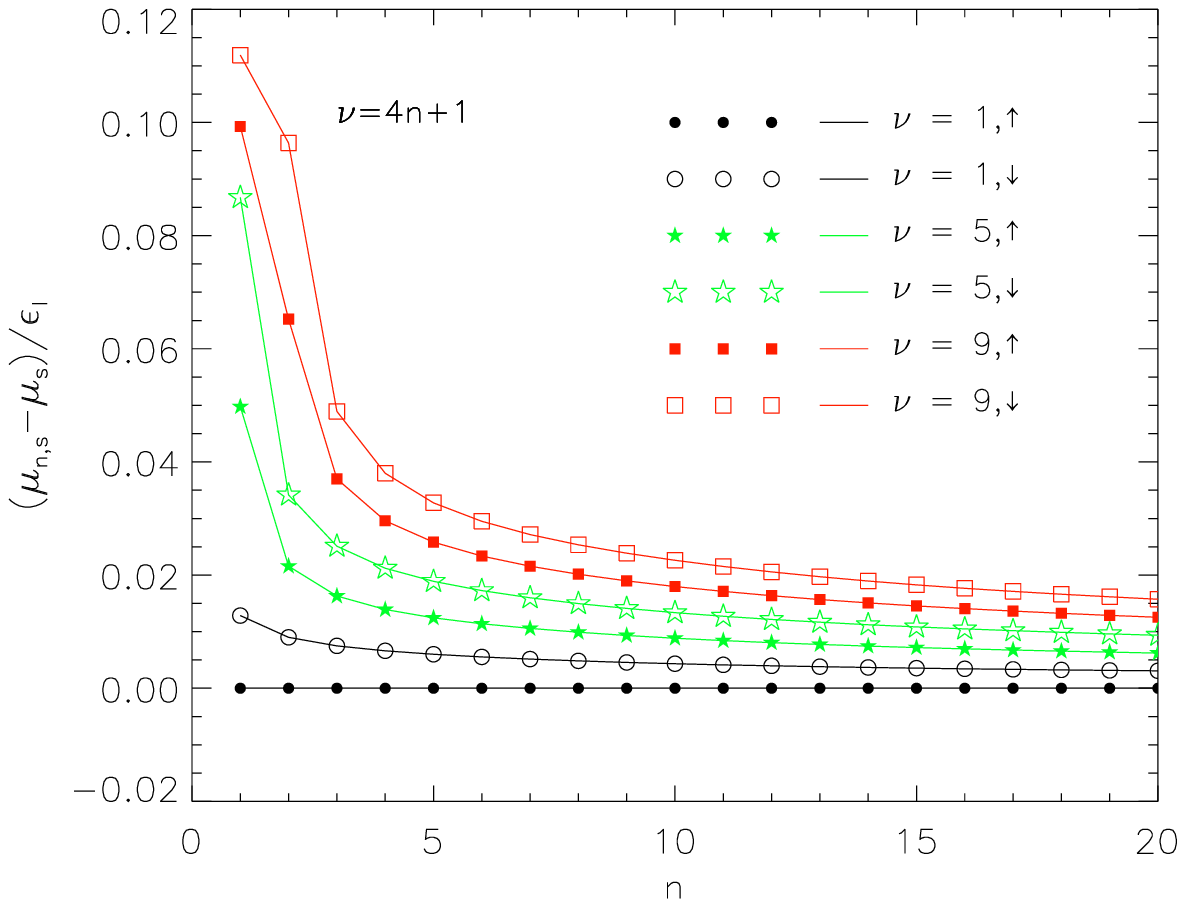}
\includegraphics[width=0.45\textwidth]{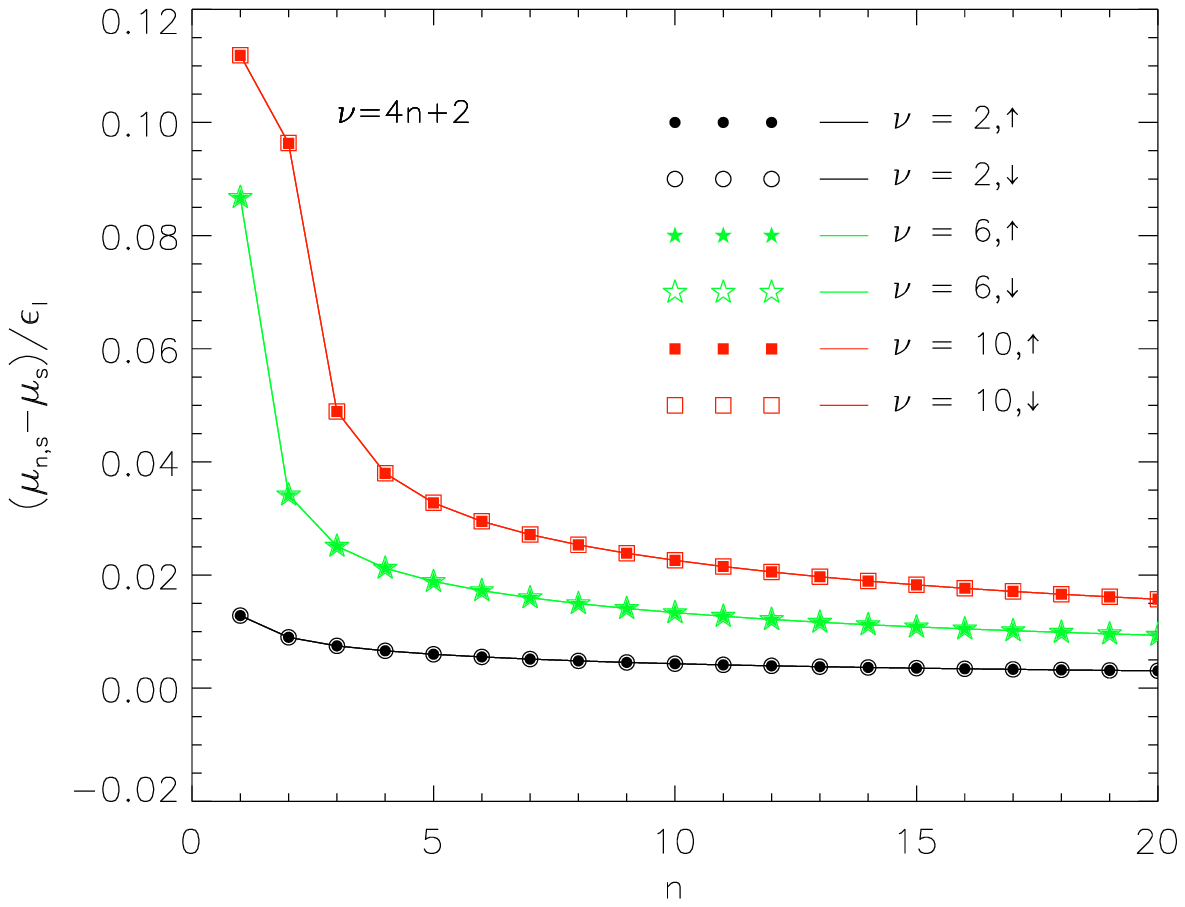}
\caption{(Color online) 
The chemical potential differences $\mu_{n,s}-\mu_s$ as functions of the Landau level index $n$ 
for four different types of QH states with filling factors $\nu=4n-1$,  $\nu=4n$,  $\nu=4n+1$, and 
$\nu=4n+2$.}
\label{fig_mu-n}
\end{figure}

\begin{figure}[ht]
\centering
\includegraphics[width=0.45\textwidth]{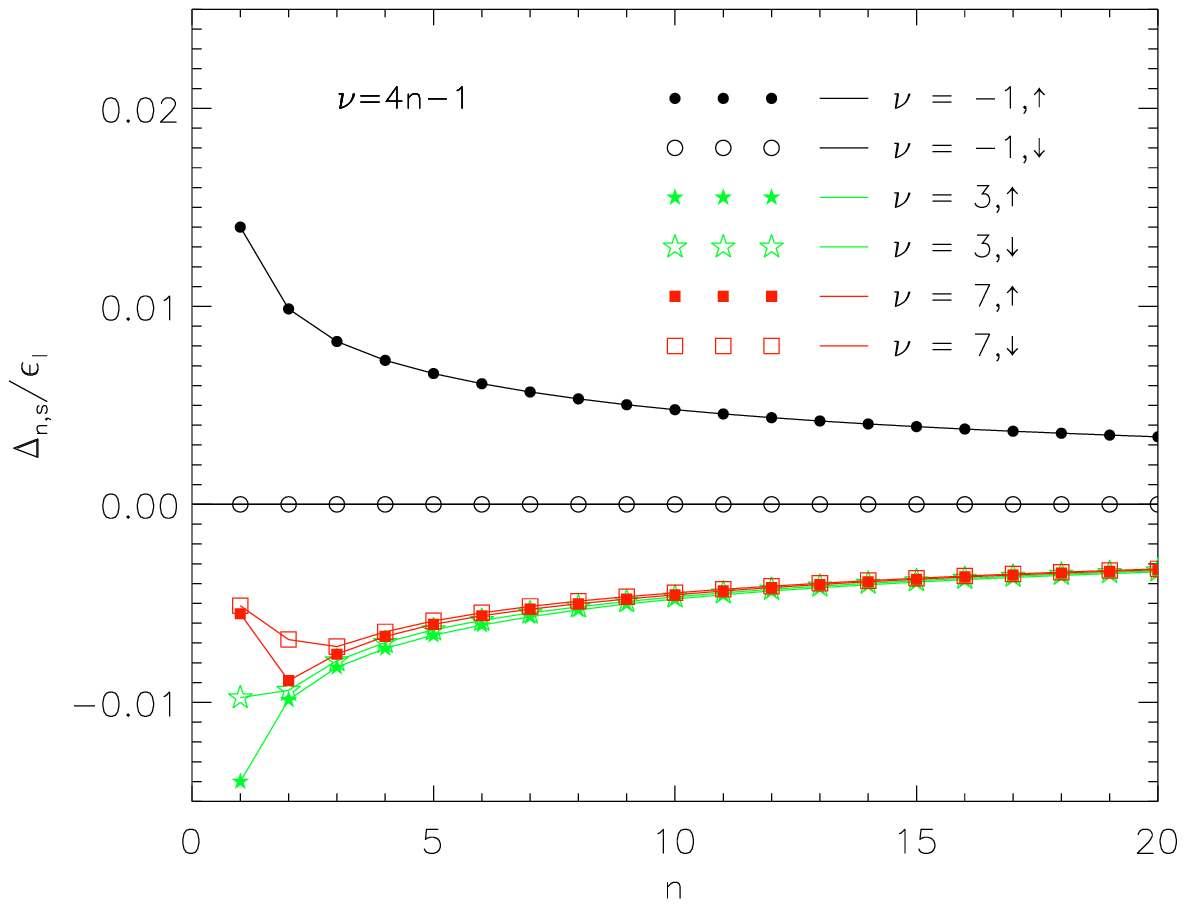}
\includegraphics[width=0.45\textwidth]{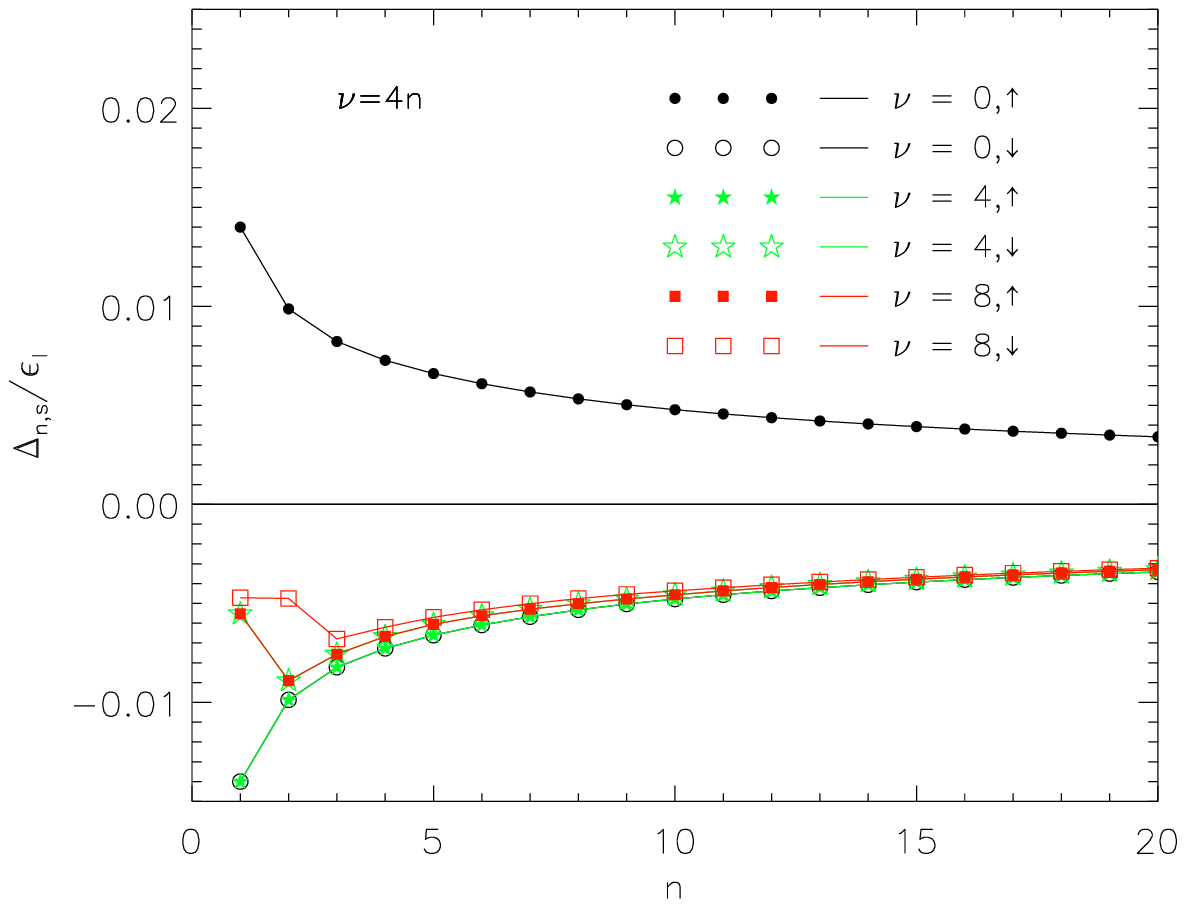}\\
\includegraphics[width=0.45\textwidth]{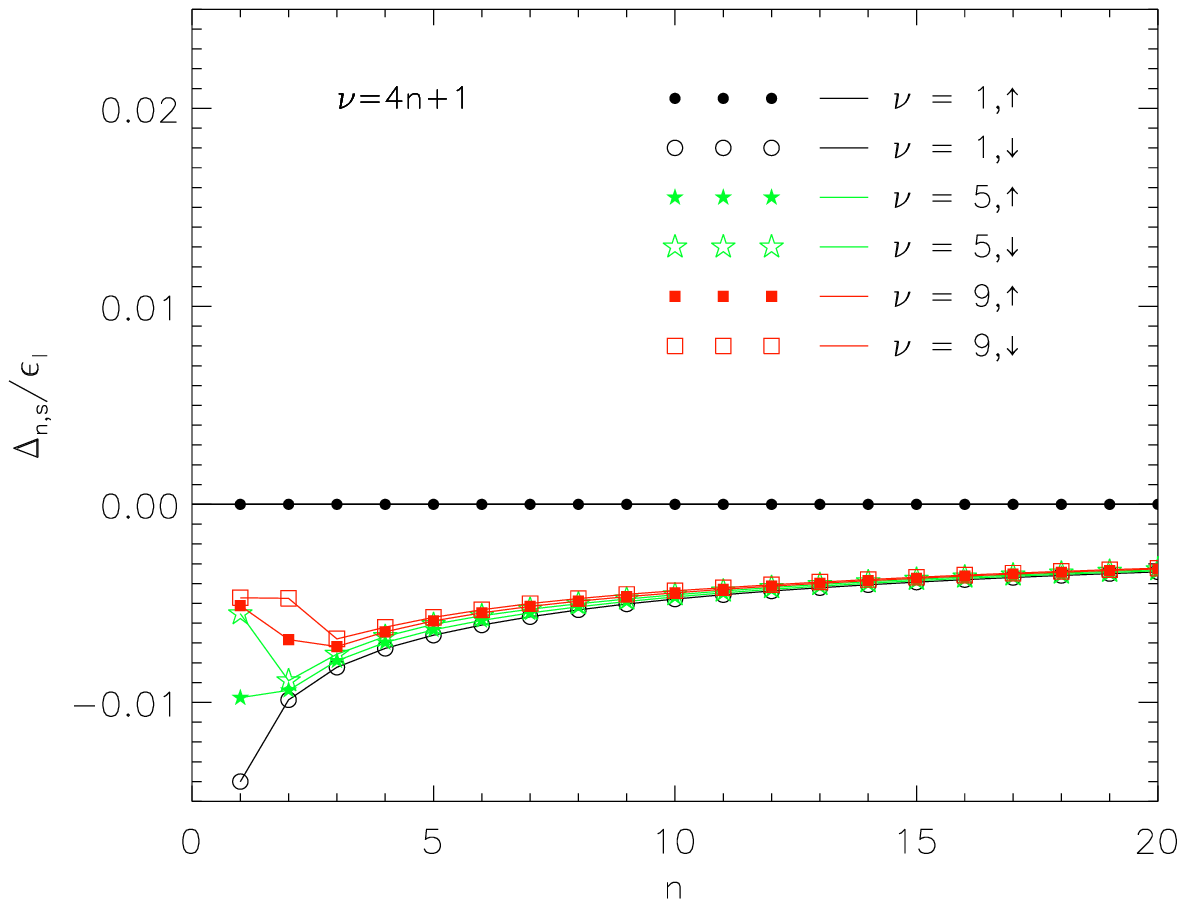}
\includegraphics[width=0.45\textwidth]{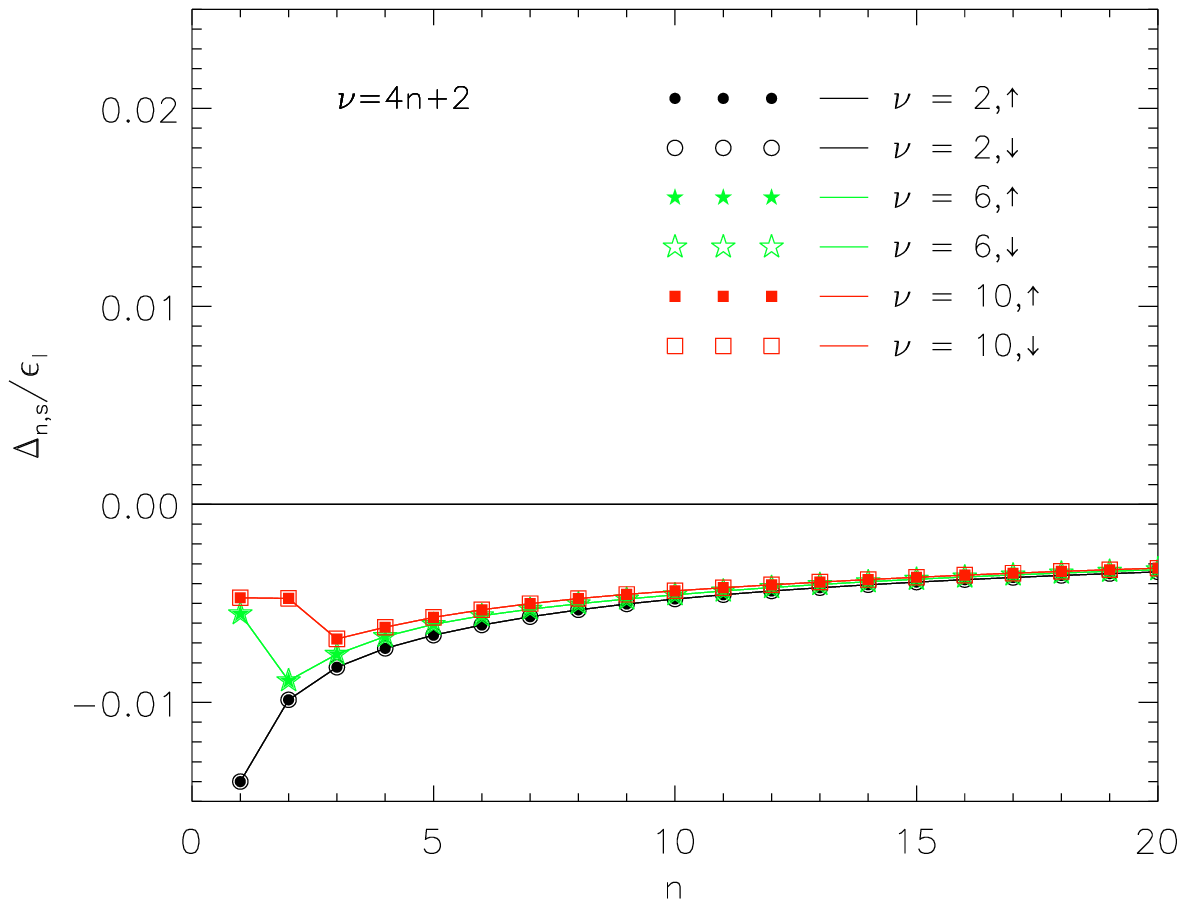}
\caption{(Color online)
The singlet Haldane masses $\Delta_{n,s}$ as functions of the Landau level index $n$ 
for four different types of QH states with filling factors $\nu=4n-1$, $\nu=4n$, $\nu=4n+1$,
and $\nu=4n+2$.}
\label{fig_Delta-n}
\end{figure}

\begin{figure}[ht]
\centering
\includegraphics[width=0.45\textwidth]{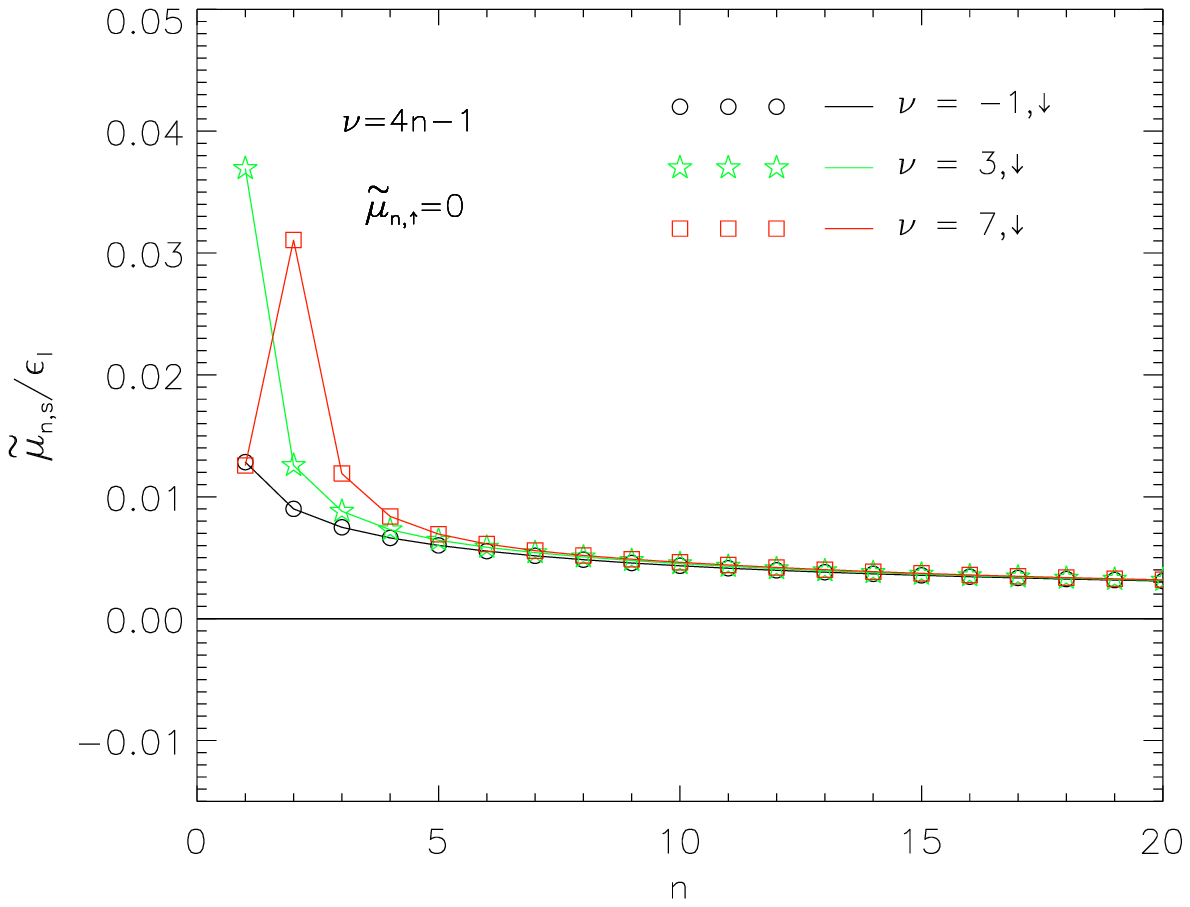}
\includegraphics[width=0.45\textwidth]{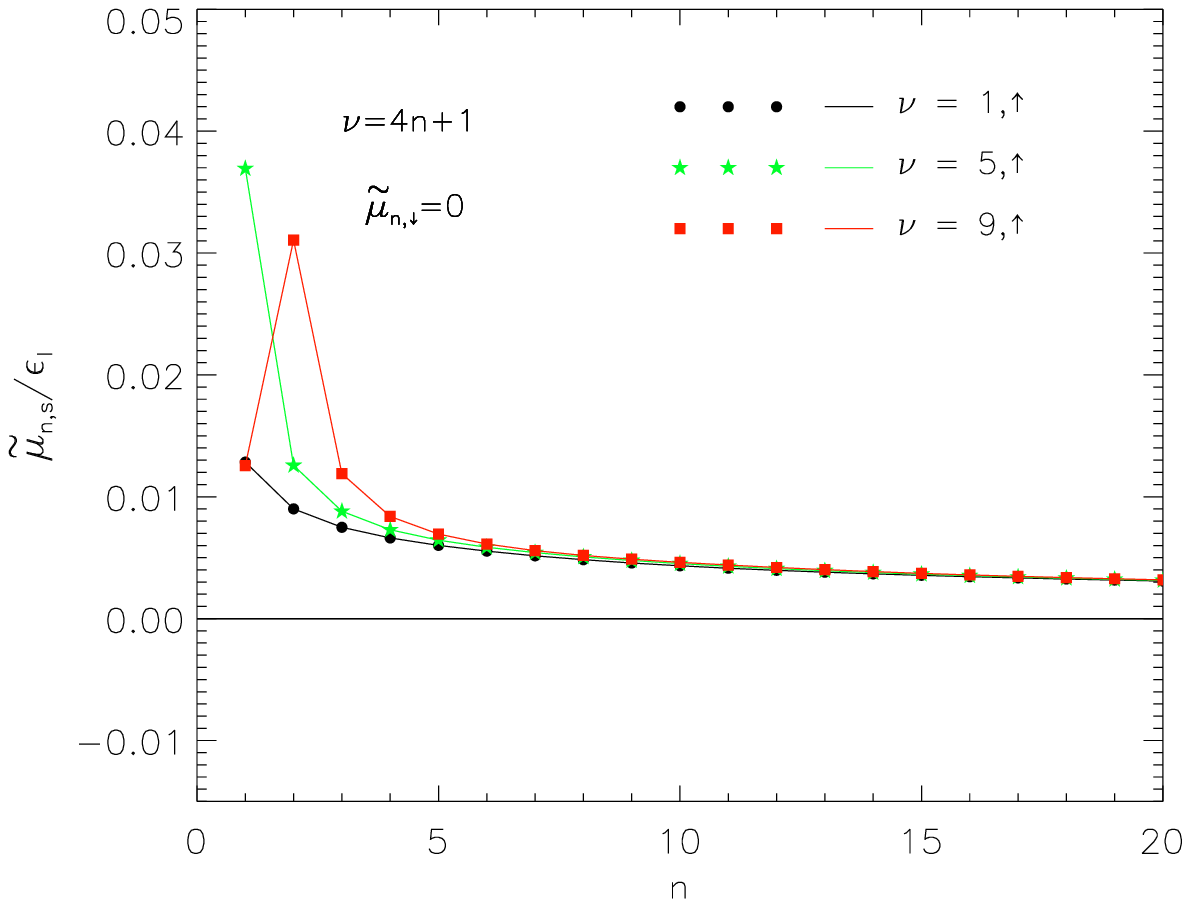}\\
\includegraphics[width=0.45\textwidth]{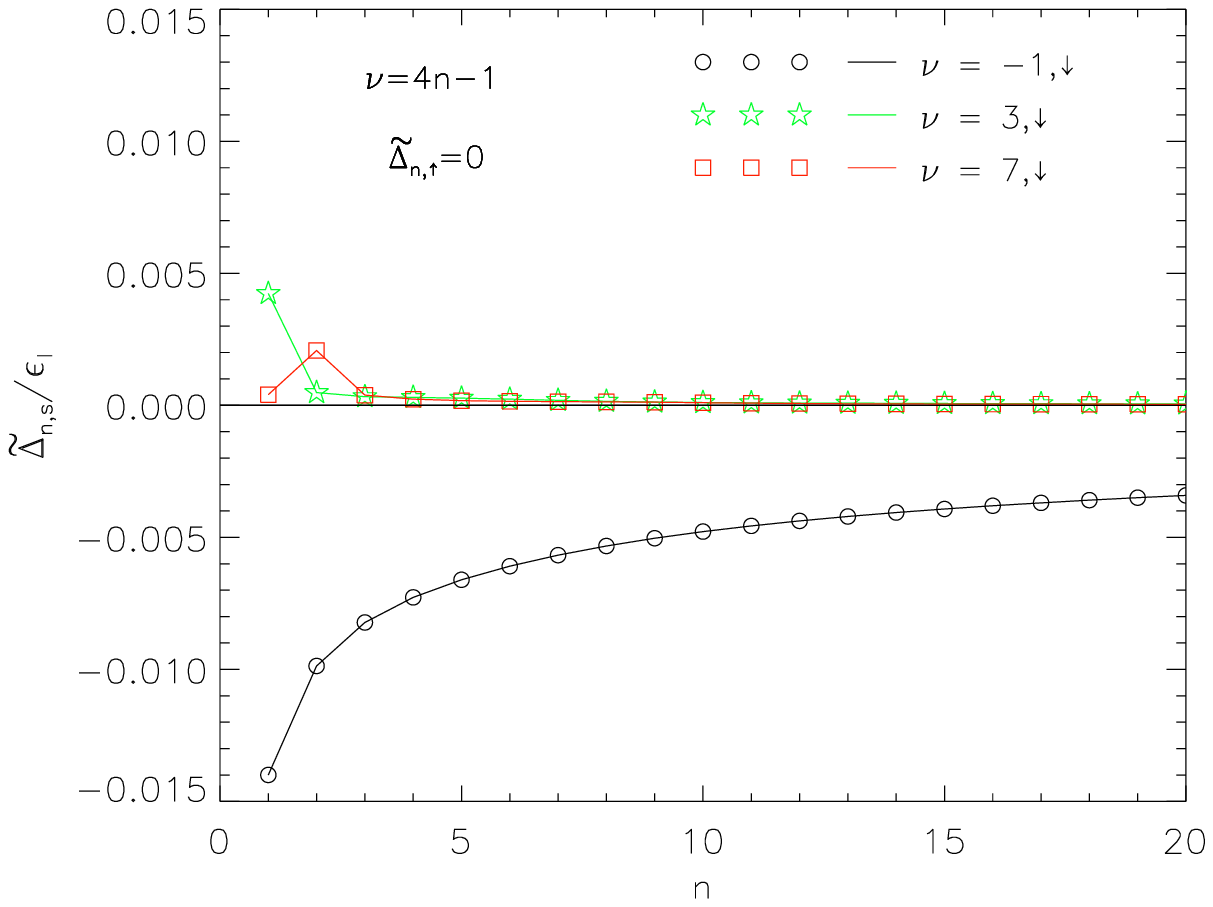}
\includegraphics[width=0.45\textwidth]{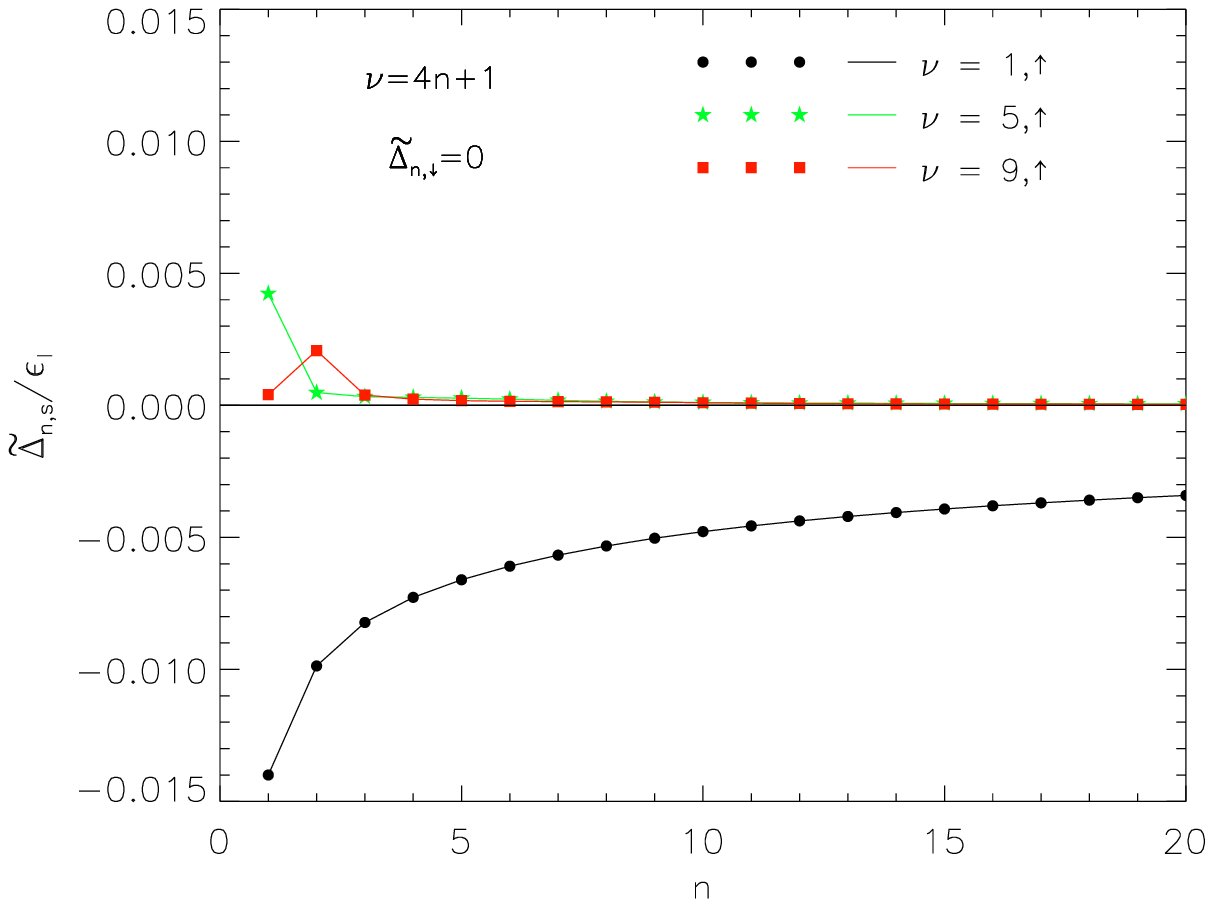}
\caption{(Color online) 
The triplet chemical potentials $\tilde{\mu}_{n,s}$ and the Dirac masses $\tilde{\Delta}_{n,s}$ 
as functions of the Landau level index $n$ for the QH states with the filling factors $\nu=4n\pm 1$.}
\label{fig_tDelta-n}
\end{figure}

Our numerical results for the running dynamical parameters $f_{n,s}$,  $\mu_{n,s}$, 
$\Delta_{n,s}$, $\tilde{\mu}_{n,s}$, and $\tilde{\Delta}_{n,s}$ are shown in Fig.~\ref{fig_f-n} 
through Fig.~\ref{fig_tDelta-n}. In line with our discussion of the four different classes of QH 
states with different filling factors, we show the results for $\nu=4n+2$, $\nu=4n$, $\nu=4n-1$, 
and $\nu=4n+1$ states in separate panels. They are characterized by different ground state 
symmetries. In order to avoid overcrowding the figures, we showed the results only for a few 
states in the series, that correspond to partially or fully filled $n=0$, $n=1$, and $n=2$ Landau 
levels. It is natural to expect that the other states in the same series, associated with filling of 
higher Landau levels, share essentially the same qualitative features. 

In Figs.~\ref{fig_f-n} -- \ref{fig_tDelta-n}, the results for the spin-up and spin-down quasiparticle 
states are represented by the same types of filled and unfilled symbols, respectively. The universal 
property of all dynamical parameters is that their values approach the free model limit at large
$n$. Additionally, we see that often the running parameters acquire their largest absolute values 
in the Landau levels near the Fermi energy. These features were expected, of course.

%%%%%%%%%%%%%%%%%%%%%%%%%%%%%%%%%%%%%%%%
%%%%%%%%%%%%%%%%%%%%%%%%%%%%%%%%%%%%%%%%

%%%%%%%%%%%%%%%%%%%%%%%%%%%%%%%%%%%%%%%%
%%%%%%%%%%%%%%%%%%%%%%%%%%%%%%%%%%%%%%%%

%%%%%%%%%%%%%%%%%%%%%%%%%%%%%%%%%%%%%%%%
%%%%%%%%%%%%%%%%%%%%%%%%%%%%%%%%%%%%%%%%

%%%%%%%%%%%%%%%%%%%%%%%%%%%%%%%%%%%%%%%%
%%%%%%%%%%%%%%%%%%%%%%%%%%%%%%%%%%%%%%%%

The results for the wave-function renormalization parameters $f_{n,s}$ as functions of $n$ are shown
in Fig.~\ref{fig_f-n}. As we see, the results are qualitatively the same for all four different classes of 
QH states. This suggests that the wave-function renormalization parameters and, thus, the renormalized
Fermi velocities are largely determined by the long-range (screened) Coulomb interaction and not 
very sensitive to the effects of the dynamically generated symmetry breaking terms. 

In the states with the even filling factors $\nu=4n$ and $\nu=4n+2$, only the singlet order parameters 
$\mu_{n,\downarrow}-\mu_{n,\uparrow}$ and $\Delta_{n,s}$ are generated. In both cases, the dynamical 
parameters have comparable magnitudes and the ground state symmetry is $U_\uparrow(2) \times 
U_\downarrow(2)$. While the role of nontrivial $\mu_{n,\downarrow}-\mu_{n,\uparrow}$ and $\Delta_{n,s}$ 
is critical in the series of states with the filling factors $\nu=4n$, it is not the case for the states with $\nu=4n+2$. 
Indeed, in the $\nu=4n$ states, it is the singlet order parameters that determine the energy gaps of the QH states. 
They are critical because the corresponding dynamical energy gaps are generated to be much larger than the 
bare Zeeman splitting. This is in contrast to the $\nu=4n+2$ states, which are characterized by rather large 
gaps (of order $\epsilon_l$) between Landau levels, which cannot be affected substantially by relatively 
small corrections due to $\mu_{n,\downarrow}-\mu_{n,\uparrow}$ and $\Delta_{n,s}$. 

Our results for the $\nu=4n$, associated with higher Landau levels, appear to be in qualitative 
agreement with the experimental results reported in Refs.~\cite{Young2012,Chiappini2015}. The 
spin-polarized nature of the corresponding states is supported by the observed increase of the activation 
gaps as functions of the in-plane component of the magnetic field.

In agreement with our general symmetry arguments, the triplet order parameters $\tilde{\mu}_{n,s}$ 
and $\tilde{\Delta}_{n,s}$ are generated only in the states with the odd filling factors $\nu=4n\pm 1$, see
Fig.~\ref{fig_tDelta-n}. We should note, however, that in both types of states the singlet order parameters 
$\mu_{n,\downarrow}-\mu_{n,\uparrow}$ and $\Delta_{n,s}$ are generated as well, see Figs.~\ref{fig_mu-n} 
and \ref{fig_Delta-n}. This is not surprising since they do not break any additional symmetries. The qualitative 
difference between the states with $\nu=4n-1$ and $\nu=4n +1$ is that the triplet order parameters are generated 
either exclusively for the spin-down quasiparticles ($\nu=4n-1$) or exclusively for the spin-up quasiparticles 
($\nu=4n+1$). It is not immediately clear whether these results are in perfect agreement with the 
experimental data in Ref.~\cite{Young2012}. Our theoretical model appears to capture some of the key 
qualitative features of the experimental data. For example, the corresponding states are characterized by
a spontaneous breakdown of the flavor symmetry and the gaps are not particularly sensitive to the Zeeman 
energy. On the other hand, the exact symmetry breaking pattern in the observed states with filling factors 
$\nu=4n\pm 1$ is not clear. In theory, the ground state symmetry is either $U_{\downarrow}(1)\times 
U_{\uparrow}(2)$ or $U_{\downarrow}(2)\times U_{\uparrow}(1)$. While this does not contradict the 
measurements, the actual symmetry could in principle be lower, i.e.,  $U_{\downarrow}(1)\times 
U_{\uparrow}(1)$. One way to resolve the issue unambiguously is to investigate the spectrum of 
quasiparticles in detail, e.g., by studying systematically all allowed optical transitions. 
We hope that this will be done in future investigations.

\subsection{Temperature dependence of the energy gaps}

As is clear from our discussion in the previous subsection, the QH states with the filling factors $\nu=4n$ and 
$\nu=4n\pm 1$ are characterized by the energy gaps that are largely determined by the dynamically generated QHF 
and MC order parameters. (Recall that the energy gaps in the $\nu=4n+2$ states are of the order of the Landau 
energy scale $\epsilon_l$ and, thus, are not affected much by the dynamical order parameters.) The corresponding 
parameters and, therefore, the energy gaps are expected to have a strong temperature dependence. In this 
subsection, we study such a dependence in detail. 

\begin{figure}[ht]
\centering
\includegraphics[width=0.45\textwidth]{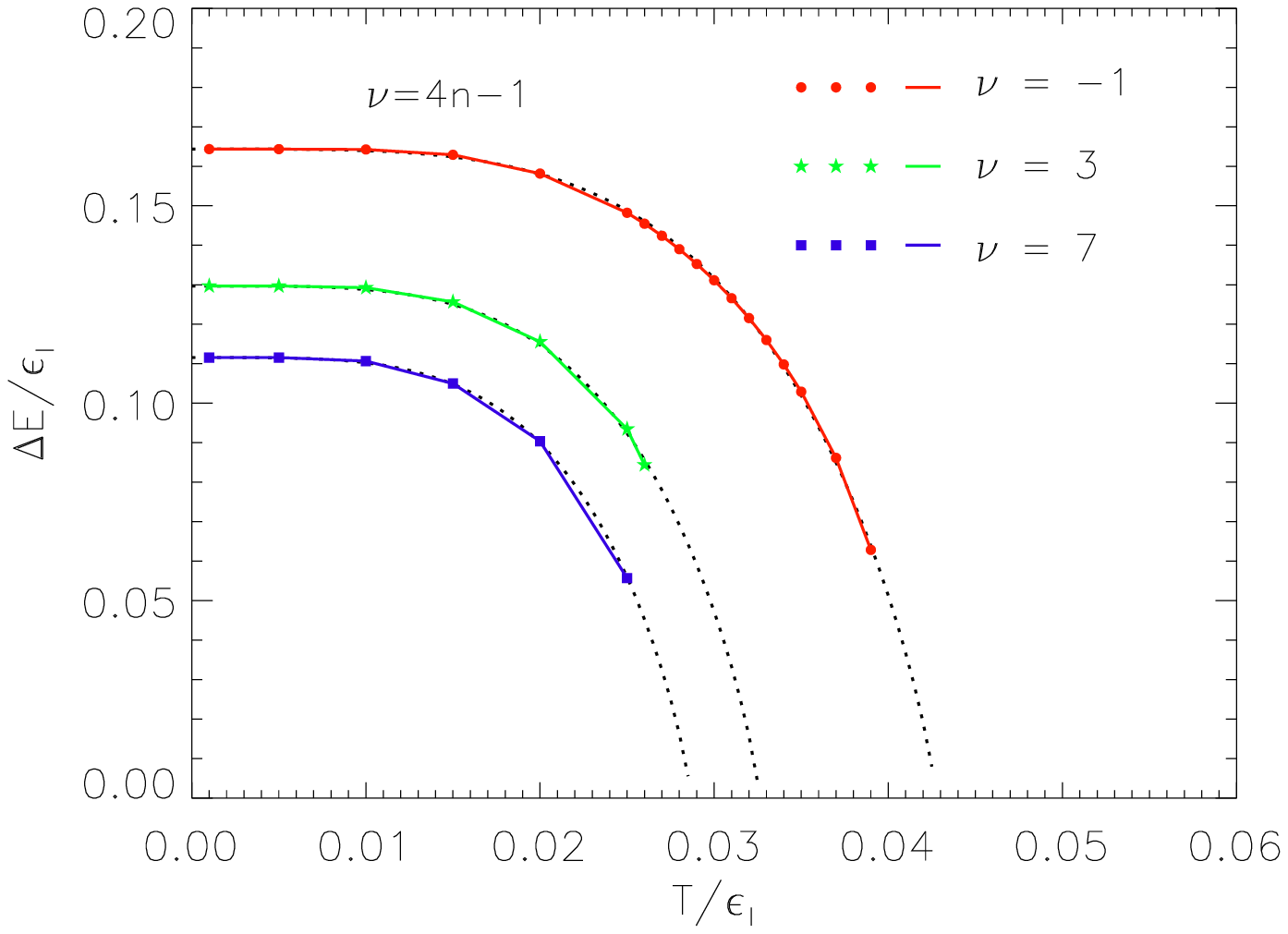}
\includegraphics[width=0.45\textwidth]{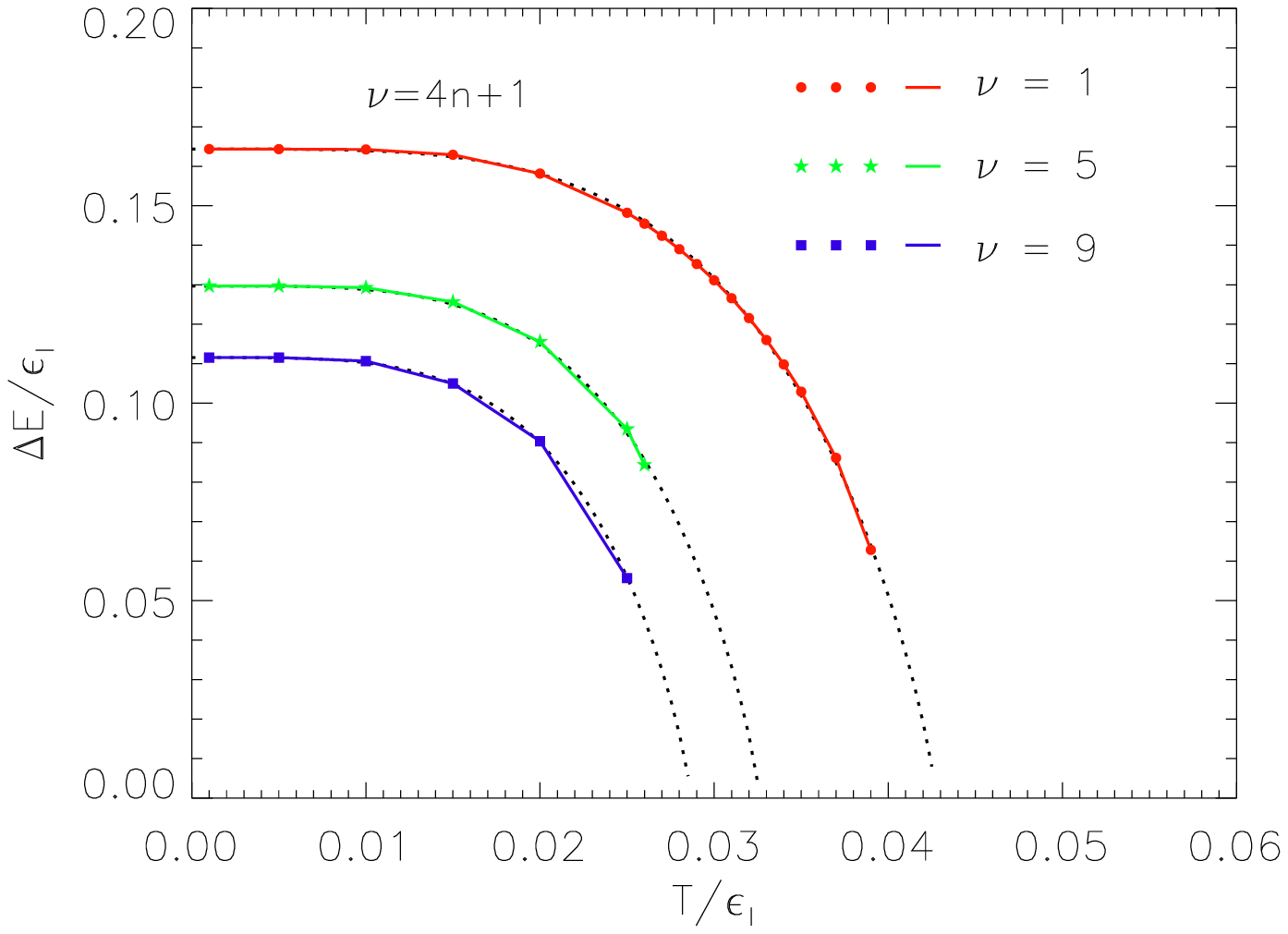}
\caption{(Color online) 
The energy gaps as functions of temperature for the QH states with the filling factors $\nu=4n\pm1$.}
\label{fig_gap-T}
\end{figure}

\begin{figure}[ht]
\centering
\includegraphics[width=0.45\textwidth]{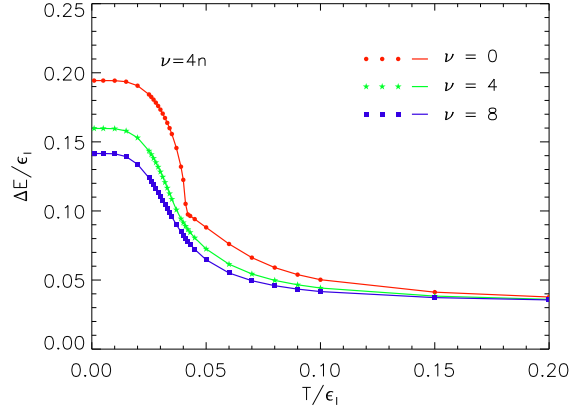}
\caption{(Color online) 
The energy gaps as functions of temperature for the QH states with the filling factors $\nu=4n$.}
\label{fig_gap_2n-T}
\end{figure}

The numerical analysis of the GLLR gap equations (\ref{gap-eq-1}) through (\ref{gap-eq-4}) at nonzero 
temperature is qualitatively the same as at $T=0$. After determining the ground states at various filling 
factors as a function of temperature, we can straightforwardly extract the temperature dependence of 
the energy gaps. For different types of the QH states, associated with various fillings of the lowest three 
Landau levels ($n=0,1,2$), our numerical results are shown in Figs.~\ref{fig_gap-T} and \ref{fig_gap_2n-T}. 
There are several universal features of these results: (i) the gaps decrease monotonically with temperature,
(ii) the overall size of the gaps decreases with increasing the Landau level index $n$. We also find that, 
for a fixed $n$, the gap functions in the states with $\nu=4n-1$ and $\nu=4n+1$ are almost exactly the 
same, see Fig.~\ref{fig_gap-T}.  

By comparing the results in Fig.~\ref{fig_gap_2n-T} with those in Fig.~\ref{fig_gap-T}, we see that the 
temperature dependence of the energy gaps in the QH states with filling factors $\nu=4n\pm 1$ is 
qualitatively different from that in the $\nu=4n$ states. By taking into account the very different 
symmetry properties of the corresponding ground states, this is not surprising at all. The $\nu=4n\pm 1$ 
states are characterized by the triplet order parameters $\tilde{\mu}_{n,s}$ and $\tilde{\Delta}_{n,s}$, 
which vanish in a symmetry restoring phase transition at the critical temperature $T_c$, see 
Fig.~\ref{fig_gap-T}. The transition appears to be either a second-order, or a weakly first-order 
transition. The temperature dependence of the energy gaps in the $\nu=4n\pm 1$ states is fitted 
quite well by the following function: 
\begin{equation}
\Delta E = \Delta E^{(0)}  \left[1-\left(\frac{T}{T_c}\right)^4\right]^{0.8},
\end{equation}
where  $\Delta E^{(0)}$ is the energy gap at zero temperature and $T_c$ is the critical temperature.  
Note that the approximate numerical values of the zero temperature gaps are $0.165\epsilon_l$, 
$0.130\epsilon_l$ and $0.112\epsilon_l$ for the $\nu=4n\pm 1$ states, associated with the Landau 
levels  $n=0,1,2$, respectively. The corresponding approximate values of the critical temperatures 
are $0.043\epsilon_l$, $0.033 \epsilon_l$ and $0.029\epsilon_l$, respectively.

In contrast, the $\nu=4n$ states are characterized by the singlet order parameters $\mu_{n,\downarrow}
-\mu_{n,\uparrow}$ and $\Delta_{n,s}$, which have the same symmetry as the Zeeman term and, thus, 
remain nonzero even at large temperatures, see Fig.~\ref{fig_gap_2n-T}. As a result, the corresponding 
transitions from the low-temperature $\nu=4n$ states with a dynamically enhanced Zeeman splitting to 
the high-temperature states without such an enhancement are generically smooth crossovers. In the 
case of the $\nu=0$ state, however, we find a sign of a small discontinuity in the temperature dependence 
of the energy gap around $T\simeq 0.04\epsilon_l$, suggesting a weak first-order transition. Of course, 
such a transition is not related to a restoration of any exact symmetry and, thus, can be viewed as 
accidental.   

%%%%%%%%%%%%%%%%%%%%%%%%%%%%%%%%%%%%%%%%
%%%%%%%%%%%%%%%%%%%%%%%%%%%%%%%%%%%%%%%%

%%%%%%%%%%%%%%%%%%%%%%%%%%%%%%%%%%%%%%%%
%%%%%%%%%%%%%%%%%%%%%%%%%%%%%%%%%%%%%%%%

\section{Discussion}
\label{sec:discussion}

In this paper, we utilized a highly flexible GLLR representation for the fermion propagator to 
describe the QH states with integer filling factors in the low-energy model of graphene with 
long-range Coulomb interaction. By including the static screening effects in an external magnetic 
field, we amended the earlier study of Ref.~\cite{gorbar:2012ps}. While using a similar 
mean-field approximation, we found that the static screening has a substantial suppression 
effect on the dynamical order parameters in all QH states with spontaneously broken 
symmetry. Also, the deviations of the wave-function renormalization from $1$ and the 
dynamical corrections to the Fermi velocity became noticeably suppressed. 

By making use of the framework that naturally incorporates the running of the dynamical 
parameters with the Landau level index $n$, we observed that the largest absolute values 
of such parameters are typically obtained for the Landau level near the Fermi energy. In 
the limit of large $n$, all dynamical parameters approach the corresponding bare values. 
By making use of numerical methods, we studied in detail the behavior of the dynamical 
parameters in all four different types of the QH states with the filling factors $\nu=4n+2$, 
$\nu=4n$, and $\nu=4n\pm 1$ that have different symmetry properties. At weak fields,
we also analyze the running of the renormalized Fermi velocity with the Landau level 
index $n$. 

In the low-energy model used, the states with the filling factors $\nu=4n+2$ and $\nu=4n$ 
have the $U_\uparrow(2)\times U_\downarrow(2)$ symmetry. They are characterized 
by the singlet type of the QHF and MC order parameters with respect to $U_s(2)$ for 
both $s=\downarrow$ and $s=\uparrow$. While the role of the corresponding singlet 
parameters in the $\nu=4n+2$ states is negligible, it is profound in the $\nu=4n$ states,
where they determine the magnitude of the dynamically enhanced Zeeman splitting.
This finding qualitatively agrees with the data reported in Refs.~\cite{Young2012,
Chiappini2015}, which supports spontaneous symmetry breaking and 
spin polarization in $\nu=4n$ states (with the exception of the $\nu=0$ state). 
The states with the odd filling factors $\nu=4n\pm 1$ 
have rather different properties and are characterized by a lower ground state symmetry, 
i.e., either $U_\uparrow(1) \times U_\downarrow(2)$ or $U_\uparrow(2)\times U_\downarrow(1)$. 
In such states, in addition to the singlet order parameters, there are two types of spontaneously 
generated triplet parameters. The latter play a critical role in the realization of the 
$\nu=4n\pm 1$ states by reducing the symmetry so as to allow the appropriate lifting 
of the fourfold degeneracy, as well as needed partial filling of Landau levels. These
properties appear to be in agreement with the experimental data in Ref.~\cite{Young2012},
even though the data was not sufficient to establish unambiguously the underlying 
symmetry breaking pattern. 

By extending the analysis to the case of nonzero temperature, we studied the energy 
gaps in the QH states with the filling factors $\nu=4n$ and $\nu=4n\pm 1$. We found 
that the symmetry breaking (triplet) order parameters, describing the $\nu=4n\pm 1$ 
states, vanish at a certain critical value of temperature $T_c$, where a second-order 
or weakly first-order transition occurs. In terms of the zero temperature gap $\Delta E^{(0)}$, 
the results for the critical temperature are approximately given by $T_c\approx 0.26 
\Delta E^{(0)}$ for all $\nu=4n\pm 1$ states associated with filling of the first few Landau 
levels. On the other hand, we do not detect a well-defined symmetry-restoring phase 
transition in the temperature dependence of the energy gaps in the $\nu=4n$ states. 
Instead, we find a smooth crossover, which is consistent with the fact that the corresponding 
states have no real symmetry-breaking order parameters. 

The results of this study unambiguously suggest that the QH states of graphene with 
various integer filling factors are generically characterized by a rather large number 
of dynamical parameters that have a nontrivial running with the Landau level index $n$. 
The corresponding rich structure of Landau levels in the QH regime could be probed 
in detail via a systematic study of transition lines in optical experiments such as those 
reported in Refs.~\cite{sadowski:2006prl,jiang:2007prl,orlita:2010sst}. The temperature
dependence of the energy gaps obtained in this study appears to be in a qualitative 
agreement with the measurements of the activation energies in Ref.~\cite{zhang:2006prl,
jiang1:2007prl,Checkelsky2008,du:2009ntr,bolotin:2009ntr,Young2012,Young2014}. 
While such an agreement is encouraging, it remains to be seen whether the theoretical 
predictions of the GLLR formalism could be matched quantitatively to the experimental 
data. Before such a comparison is attempted, however, one may need to reevaluate the
model approximations used in the present study. In particular, one should address the 
precise quantitative role of (i) dynamical screening, (ii) fluctuations of the order parameter 
fluctuations, and (iii) nonzero quasiparticles widths. Hopefully, all such effects in GLLR 
formalism will be addressed in the future studies.

\acknowledgments
The authors acknowledge participation of Yingchao Lu during the early stages of this project. 
The authors thank E. V. Gorbar for numerous discussions and useful comments on the text of the paper.
This work was supported by the U.S. National Science Foundation under Grant No.~PHY-1404232.

\end{document}